\documentclass[useAMS,usenatbib,onecolumn]{mn2e}
\usepackage{graphicx}
\usepackage{lscape}
\usepackage{longtable}
\usepackage{array}
\newcommand{\oversim}[2]{\protect{\mbox{\lower0.5ex\vbox{%
  \baselineskip=0pt\lineskip=0.2ex
  \ialign{$\mathsurround=0pt #1\hfil##\hfil$\crcr#2\crcr\sim\crcr}}}}}
\newcommand{\simgreat}{\mbox{$\,\mathrel{\mathpalette\oversim>}\,$}} 
\newcommand{\simless} {\mbox{$\,\mathrel{\mathpalette\oversim<}\,$}} 
\title[Star formation and massive stars]{The relation between the 
  most-massive star and its parental star cluster mass}
\author[C.~Weidner, P.~Kroupa and
  I.~A.~D.~Bonnell]{C.~Weidner$^{1,2}$\thanks{E-mail:
    cw60@st-andrews.ac.uk}, P.~Kroupa$^{3}$\thanks{E-mail:
    pavel@astro.uni-bonn.de} and I.~A.~D.~Bonnell$^{1}$\thanks{E-mail: 
    iab1@st-andrews.ac.uk}\\
$^{1}$Scottish Universities Physics Alliance (SUPA), School of Physics and
  Astronomy, University of St. Andrews, North Haugh,\\
 St. Andrews, Fife KY16 9SS, UK\\
$^{2}$Departamento de Astronom{\'i}a y Astrof{\'i}sica, Pontificia
      Universidad Cat{\'o}lica de Chile, Av. Vicu{\~n}a MacKenna 4860,\\
      Macul, Santiago, Chile\\
$^{3}$Argelander-Institut f\"ur Astronomie (Sternwarte), Auf dem H{\"u}gel 71,
D-53121 Bonn, Germany
}

\begin{document}
\bibliographystyle{aa}
\date{Accepted . Received ; in original form }

\pagerange{\pageref{firstpage}--\pageref{lastpage}} \pubyear{2009}

\maketitle

\label{firstpage}

\begin{abstract}
We present a thorough literature study of the most-massive star,
$m_\mathrm{max}$, in several young star clusters in order to assess
whether or not star clusters are populated from the stellar initial
mass function (IMF) by random sampling over the mass range
$0.01~\le~m~\le~150~M_\odot$ without being constrained by the cluster mass,
$M_\mathrm{ecl}$. The data reveal a partition of the sample into
lowest mass objects ($M_\mathrm{ecl} \le 10^2 M_\odot$), moderate mass
clusters ($10^2 M_\odot < M_\mathrm{ecl} \le 10^3 M_\odot$) and rich
clusters above $10^3 M_\odot$. Additionally, there is a plateau of a
constant maximal star mass ($m_\mathrm{max} \approx$ 25 $M_\odot$) for
clusters with masses between $10^3 M_\odot$ and $4 \cdot 10^3 M_\odot$. 
Statistical tests of this data set reveal that the hypothesis of
random sampling from the IMF between 0.01 and 150~$M_\odot$ is highly
unlikely for star clusters more massive than $10^2 M_\odot$ with a
probability of $p \approx 2 \cdot 10^{-7}$ for the objects with
$M_\mathrm{ecl}$ between $10^2 M_\odot$ and $10^3 M_\odot$ and
$p \approx 3 \cdot 10^{-9}$ for the more massive star clusters. Also,
the spread of $m_\mathrm{max}$ values at a given $M_\mathrm{ecl}$ is
smaller than expected from random sampling. We suggest that the basic
physical process able to explain this dependence of stellar inventory
of a star cluster on its mass may be the interplay between stellar
feedback and the binding energy of the cluster-forming molecular cloud
core. Given these results, it would follow that an integrated galactic
initial mass function (IGIMF) sampled from such clusters would
automatically be steeper in comparison to the IMF within individual
star clusters.
\end{abstract}


\section{Introduction}
\label{se:intro}

\begin{figure}
\begin{center}
\includegraphics[width=8cm]{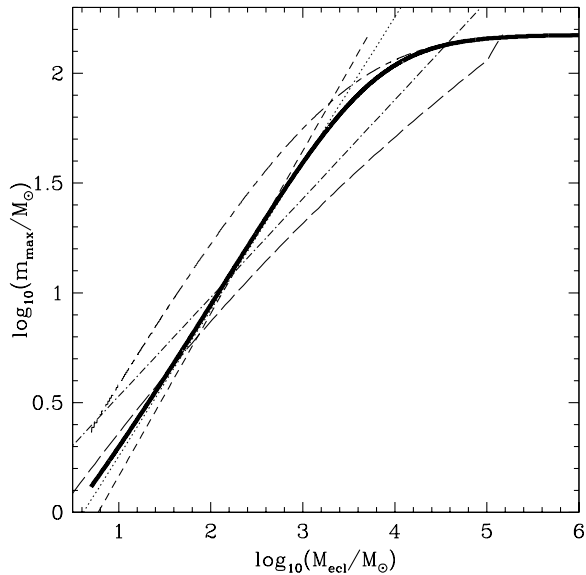}
\vspace*{-2.0cm}
\caption{Different relations between the most-massive star and the
  cluster mass from observations, numerical calculations and
  theoretical modelling from the literature. The {\it dash-dotted}
  line marks the empirical relation by \citet{La03}, the {\it
    long-dashed line} is the \citet{El83} relation, the {\it
    short-dashed line} is the \citet{Elme00b} relation, the {\it
    dotted line} shows the \citet{BBV03} relation, the {\it
    short-dashed-long-dashed line} is the \citet{OC05} relation and
    the {\it thick-solid line} is the analytical model of
    \citet{WK04}. See text for more details.}
\label{fig:mvM_hist}
\end{center}
\end{figure}

Whether or not newborn stars in star clusters are randomly
drawn\footnote{Random sampling means choosing a number $N$ of stars
  randomly from the distribution function which is in this case the IMF.} from
the IMF is of utmost importance for various fields of stellar
astrophysics. For example non-random drawing which suppresses the
number of OB stars in smaller clusters would steepen the IMF for whole
galaxies, the integrated galactic stellar initial mass function
\citep[IGIMF, ][]{KW03,WK05a}. A randomly-drawn IMF on the other hand,
which would be equivalent to postulating the existence of clusters
comprised of a massive star and not much more, would not
\citep{Em06,SM08}. As the bulk of the galactic field star populations
are probably made from dissolving star clusters
\citep{Kr95c,LL95,LL03,AM01,AMG07}, understanding the stellar distribution
in galaxies presupposes knowledge of the IMF in star clusters. A
central issue on deciding whether a star cluster can be modelled in
terms of random sampling from the IMF or not is the existence of a non-trivial
relation between the mass of the most-massive star
($m_\mathrm{max}$) and the star cluster mass ($M_\mathrm{ecl}$).
Thus, a statistically significant correlation
$m_\mathrm{max}(M_\mathrm{ecl})$ would imply physical processes such
as self-regulation of the star-formation process in a cluster. The
influence of the cluster mass or density on its 
stellar population has been studied on previous occasions. \citet{La82}
and \citet{La03} examined the properties of molecular clouds and the
stellar populations found within them, finding the following empirical
expression between the mass of the most-massive star,
$m_\mathrm{max}$, and the stellar mass of an embedded cluster, 
$M_\mathrm{ecl}$, 
\begin{equation}
m_\mathrm{max}^\mathrm{Larson} = 1.2 M_\mathrm{ecl}^{0.45},
\end{equation}
which is shown as a {\it dash-dotted line} in Fig.~\ref{fig:mvM_hist}.

\citet{El83} investigated a model for the formation of
bound star clusters where the luminosity of the stars chosen from a
\citet{MS79} IMF overcomes the binding energy of a molecular
cloud. Different star formation efficiencies would then determine if
a cloud becomes a bound star cluster or an OB association. He also
found a relation between $m_\mathrm{max}$ and $M_\mathrm{ecl}$ which
cannot be written as an analytical equation and is shown as a {\it
long-dashed line} in Fig.~\ref{fig:mvM_hist}. Later, 
\citet{Elme00b} derived a different relation when assuming a single
slope power-law IMF, $\xi(m)$, where $dN = \xi(m) dm$ is the number of
stars in the mass interval $m,~m+dm$, with a \citet{Sal55} slope and solving
the following two equations,
\begin{equation}
\label{eq:norm}
1 = \int_{m_\mathrm{max}}^{m_\mathrm{max *}} \xi(m) dm
\end{equation}
and
\begin{equation}
\label{eq:Mecl}
M_\mathrm{ecl} = \int_{m_\mathrm{min}}^{m_\mathrm{max}} m \xi(m) dm,
\end{equation}
but without any limit for masses of the stars, $m_\mathrm{max *} =
\infty$. Here, $m_\mathrm{min}$ is the minimum mass. For a single power-law 
IMF with a \citet{Sal55} slope these two equations yield in,
\begin{equation}
m_\mathrm{max}^\mathrm{Elmegreen} = \left(\frac{M_\mathrm{ecl}}{3
  \times 10^3}\right)^{-1.35} \times 100,
\end{equation}
shown as the {\it short-dashed line} in Fig.~\ref{fig:mvM_hist}.

In their numerical calculations of star-forming molecular clouds using
a smoothed particle hydrodynamics code \citet{BBV03,BVB04} found a relation,
\begin{equation}
m_\mathrm{max}^\mathrm{Bonnell} = 0.39 \times M_\mathrm{ecl}^{2/3},
\end{equation}
which is shown as a {\it dotted line} in Fig.~\ref{fig:mvM_hist}.

In a thorough study of star clusters and OB associations in order to
determine whether or not a fundamental upper mass limit for stars
exists, \citet{OC05} also calculated the expected dependence of
$m_\mathrm{max}$ on $M_\mathrm{ecl}$ if the stars are randomly drawn
from an IMF,
\begin{equation}
m_\mathrm{max}^\mathrm{Oey} = m_\mathrm{max *} - \int_0^{m_\mathrm{max
    *}} \left[ \int_0^{M_\mathrm{ecl}} \xi(m) dm \right]^N dM_\mathrm{ecl},
\end{equation}
plotted as a {\it short-dashed-long-dashed line} in
Fig.~\ref{fig:mvM_hist}.

Including a fundamental upper mass limit for stars, $m_\mathrm{max *}
= 150 M_\odot$\footnote{150 $M_\odot$ is believed to be the fundamental
upper mass limit for stars with non-zero metallicity
\citep{WK04,OC05,Fi05,Ko06}.} in eqs.~\ref{eq:norm} and \ref{eq:Mecl},
and using the canonical multi-part power-law IMF (Appendix~\ref{app:IMF})
\citet{WK04} found the relation visible as a {\it thick-solid line}
in Fig.~\ref{fig:mvM_hist}. 

As evident from Fig.~\ref{fig:mvM_hist} these studies arrive at a
rather large range of possible
$m_\mathrm{max}$--$M_\mathrm{ecl}$-relations. \citet{WK05b}
re-investigated this question by compiling a larger number of
observational results from the literature and extensive Monte-Carlo
experiments of different sampling algorithms and found evidence that
there exists a non-trivial relation between the 
mass of a star cluster and the most-massive star in the cluster, 
a result in principle confirmed by \citet{SM08}. But they
  conclude that the \citet{WK05b} sample is biased by a size-of-sample
effect. Furthermore, in the recent literature several claims have been made 
against such a relation arguing instead for a pure random sampling from
the IMF in individual star clusters \citep{OKP04,DTP05,Em06,PG07,SM08,MC08}. 
\citet{DTP04,DTP05} find up-to 4\% of non-runaway (less
than 30 km/s space motion) O stars in isolation
with no apparent cluster around them or within their lifetime if they
would have been ejected from a cluster with a velocity of 6 km/s --
indicating they formed outside a cluster. This result would of
course be irreconcilable with a relation between the mass of the
most-massive star and the mass of its parent star cluster as has been
pointed out by \citet{PG07} and \citet{SM08}. While \citet{SM08} argue
that the sample used in \citet{WK05b} is biased against random
sampling, \citet{PG07} find that the observed 4\% of allegedly
isolated O stars would agree with 
random sampling from a cluster number distribution function which
scales with $N^2$, where $N$ is the number of stars in a cluster. But
it should be noted here that the \citet{DTP05} result is an {\it upper
limit} for O stars formed in isolation and that more in-depth
observations might reduce this sample. For example, HD 165319, 
an O9.5~I star from the \citet{DTP05} sample of 11 stars which are
indicated there as one of ``the best examples for isolated Galactic
high-mass star formation'' has a bow-shock front and is therefore a
star ejected from a star cluster, possibly NGC 6611
\citep{GB08}. Additionally, according to \citet{SR08} further 6 of the
remaining 10 stars are at distances to star clusters only slightly larger
than what they may have travelled during their expected life
times. But the current large errors of the space motion of these stars
does not allow to constrain the birth places of them.

A non-trivial $m_\mathrm{max}$-$M_\mathrm{ecl}$-relation, and
therefore whether or not the stars in star clusters are
  randomly sampled from the IMF, would also give more insight and
understanding of the process 
of star-formation. The formation of massive stars
($>$ 10 $M_\odot$) is still not well understood with at least two
competing theories (competitive accretion vs. single star accretion)
having been developed \citep{BBZ98,BVB04,BB06,TKM06,KKM09}. An
$m_\mathrm{max}$-$M_\mathrm{ecl}$-relation could imply that the bulk
of the low-mass stars form first and the high-mass stars later. The
combined feedback of the massive stars would then halt further
star-formation.\\

Here we will show that the observed distribution of the mass, 
$m_{\rm max}$, of the most-massive star in a star cluster cannot
be drawn randomly from the IMF for clusters more massive than 100
$M_\odot$ but that there must exist a physical relation between
$m_{\rm max}$ and the birth stellar mass of the cluster, $M_{\rm ecl}$
(the stellar content before gas expulsion but after cessation of star
formation).

\section{The data}
\label{se:data}
\subsection{Sample construction}
In order to construct a sufficiently large observational sample to
test whether random sampling from the IMF is an acceptable model in
star clusters the available 
literature was searched for star clusters which are young enough to
not have experienced supernova events and are dynamically rather
un-evolved. For the latter the star clusters should still be embedded
in their natal gas cloud or at least be very young, such that gas
expulsion would not have effected them strongly 
\citep{LMD84,Go97,KAH01,GB06,WKNS06,PMH06b,BG06,WLB07}. Therefore only
clusters younger than 4 Myr have been included in our sample.
Additionally, the young age limits the amount of mass-loss
experienced by massive stars due to stellar evolution.

An allegedly suitable sample of objects discussed by \citet{PG07} and
\citet{MC08} is the one compiled by \citet{TPP97,TPN98,TPN99} as these
authors were explicitly searching for clusters around young A and B
stars. We do not use the majority of the clusters from these studies
for the following reasons: a) the majority are too old ($>$~4 Myr for 25 of
35 objects) or they are b) gas-free. The age limit imposed here is given by
the short life time of massive stars and to limit stellar
  mass-loss of the massive stars. Completely gas-free objects are
unsuited for the task of this work as gas-expulsion will remove large 
amounts of stars and therefore reduce the mass of the cluster, $M_{\rm
ecl}$, significantly \citep{KAH01,WKNS06}. The exception are four
objects which this sample has in common with the near infra-red study
of young star-forming regions by \citet{WL07} and which are all
included in our study. Based on similar arguments \citet{MC08}
  also excluded the 
  \citet{TPP97,TPN98,TPN99} sample from their final statistical
  analysis.

A very recent additional sample is provided by \citet{FMT09}. The
authors study 26 high-luminosity IRAS sources and find that 22 of them
show evidence for clustering. They model 9 of these clusters in order
to derive cluster masses and the mass of the most massive stars. This
sample is included in our study, too. But because the results are
based on modelling, different symbols for them are used in subsequent
plots. \citet{FMT09} conclude that the masses of the most-massive star
in these clusters are also not reconcilable with random sampling of the
stars from the IMF.

\subsection{Mass of a cluster versus number of stars in a cluster}
\label{sub:MvsN}
The claim has been made \citep{PG07,MC08} that the number of stars
within a star cluster, 
$N_\mathrm{ecl}$, gives a better statistical description of the
cluster compared with the cluster mass, $M_{\rm ecl}$, because
$N_\mathrm{ecl}$ is an observed quantity and statistically more easily
manageable. This is, however, not entirely true as observational
biases handicap $N_\mathrm{ecl}$ to a larger extent than
$M_\mathrm{ecl}$. As the lower mass limit of the
observations depend on telescope time, distance of the object,
reddening and observed colour range, the different clusters have to be
normalised to the same lower mass limit in order to make them
comparable. This is done for $N_\mathrm{ecl}$ in 
the same way as for $M_{\rm ecl}$ -- by extrapolating the stars in the
observed mass range to a general mass range (0.01 to 150 $M_\odot$ in this
study) with the use of an IMF. Therefore, $N_\mathrm{ecl}$ is {\it
  not} an observed quantity but an estimated one. But the sources for
potential error are much larger in the case of $N_\mathrm{ecl}$ than
compared with $M_\mathrm{ecl}$, as the observed number of stars gives
every star the same statistical weight, regardless if it is an M dwarf
or an O supergiant. But low-mass stars and brown dwarfs are easy to
miss due to being faint but also due to un-resolved binarity and
crowding of stars \citep{MA08,WK07c}. 
Very young low-mass PMS stars and brown dwarfs are still difficult to
model because they are dominated by the unknown accretion history, and
magnetic fields and fast rotation have a strong influence
\citep{CGB07,RMJ08}. Therefore is the observational lower mass limit 
highly model-dependent and has large errors. Because the IMF is dominated in
number by low-mass objects (85\% of all stars are below 0.5 $M_\odot$
for the IMF described in Appendix~\ref{app:IMF}) uncertainties in the
lower mass limit severely affect the $N_\mathrm{ecl}$ estimate. The mass,
$M_\mathrm{ecl}$, in contrast is far easier to estimate by the number
of high-mass stars \citep{MA09}. Likewise, the stellar evolution
models of massive stars still include large uncertainties. As in the 
case of low-mass stars, the effects of fast rotation and magnetic
fields in these stars are not well understood. Massive stars
are small in numbers (6\% of all 
stars are above 1 $M_\odot$) but dominate the cluster in mass (50.7\%
of the total mass is in stars above 1 $M_\odot$ for a cluster
comprised of 0.01 to 150 $M_\odot$ objects according to the canonical
IMF as described in Appendix~\ref{app:IMF}). Because of the
intrinsic brightness of these objects they are easy to access
observationally and difficult to miss. While the binary frequency
might be lower for low-mass stars ($\sim$ 35\%) compared to massive stars 
\citep[20\% to 80\%,][]{GCM80,GM01,DME06,KKK06,Lu06,ABK07,SGN07,TBR08,WK07c},
the effect of un-resolved binaries is smaller for the
mass estimate than for the number estimate. If all stars were in
un-resolved binaries, $N_\mathrm{ecl}$ would miss 50\% of the stars
while $M_\mathrm{ecl}$ would miss only 16 to 30\%, depending on the
mass-ratio distribution \citep{WK07c}. We therefore choose to study
$m_\mathrm{max}$ in dependence of $M_\mathrm{ecl}$ rather than
$N_\mathrm{ecl}$.

\subsection{Additional issues}
If gas-expulsion already starts early on, before the explosion of
supernovae, even the young objects presented here might be effected by
mass-loss due to the unbinding of stars from the cluster.

One possible additional effect which might deplete very young star
clusters especially from massive stars are dynamical ejections after
stellar encounters in the dense decoupled cores of massive clusters
\citep{CP92,PAK06}. Unfortunately, this effect is impossible to avoid or to
correct for reliably and might lead to an additional underestimation
of the cluster masses, but it is unlikely that the most-massive star is
ejected from the cluster.

Already in \citet{WK05b} a first set of young star clusters and their
most massive stars were presented. In the current contribution the
\citet{WK05b} list is included, corrected for a few errors and
significantly expanded. The sample of 100 both new and previously
published star clusters is shown as Tab.~\ref{tab:clustersold} in
Appendix~\ref{app:data}. The table shows two mass values for the
  mass of the most-massive star. The one in column \# 3 is based on
  the \citet{VGS96} spectral class to stellar mass conversion. In
  column \# 4, additionally, a new spectral class to mass conversion
  is used. It is based on \citet{MSH05} and \citet{MP06} who provide
  two new transformations of O-star spectral 
types into masses, which are both rather similar. One is based on a
theoretical effective temperature scale and the other on an observational
one. The authors note that their new calibration should represent a
significant improvement over previous calibrations, due to the
detailed treatment of non-LTE line-blanketing in their
calculations. Using the new transformation based on the theoretical
effective temperature scale \citep[table~1 in][]{MSH05}, all the
clusters with O stars ($m \simgreat 16 M_{\odot}$) in
Tab.~\ref{tab:clustersold} are re-examined. The resulting new spectral
masses are corrected for stellar evolutionary effects \citep{W09b}
and the new masses for the most-massive stars are compiled in column
\# 4 of the same table. The difference between the old and the new
calibration is visiualised in Fig.~\ref{fig:old_vs_new} in
Appendix~\ref{app:data}. As is shown there, in all but four cases 
the new calibration results in stellar masses significantly lower than
the old values. Note that the new calibration by
\citet{MSH05} is only provided up to a spectral type of O3. 
This might not include the most massive stars observed but no general
consensus exists in spectral classifying of extremely massive
stars. While, traditionally they would be of spectral type O3 some
classify them as spectral type O2 or even earlier \citep{WHL02} while
others prefere a Wolf-Rayet star classification \citep[for example
WN6h,][]{Cr07}. In these cases the corrected values based on the
\citet{Cr07} Wolf-Rayet scheme in \citet{W09b} are used.

An additional complication in the determination of the masses of the
most-massive stars is due to possible binary stellar
evolution (BSE). Massive stars are often found in close binaries
of rather similiar masses \citep{WK07c} and therefore BSE might have
affected the evolution and hence the observational parameters of the stars
\citep{WDT96,TAP97,Hu03,ZHL05}.

Except for a few cases the cluster masses are derived by extrapolating
from the given number of stars above a certain mass limit or within
certain limits to a mass range of 0.01 to 150.0 $M_\odot$ with a
canonical IMF (see Appendix~\ref{app:IMF}).\\

Several cluster masses given in \citet{CSS93} are used as lower limits
only in this study because of incompleteness and uncertain
differential reddening.\\

In Appendix~\ref{app:notes} notes on some individual clusters can be found.

\subsection{Dynamical masses}
\label{sub:dyn}

In recent years observational techniques allowed to measure masses of very
massive stars directly by observing the orbits of massive eclipsing
binaries. In Table~\ref{tab:dyn} the dynamical mass estimates for six
very massive stars are compared with old and new spectroscopic
estimates. Two of these six stars (WR20a A and $\Theta$ Orionis C1)
happen to be the most massive stars in two clusters (Westerlund 2 and
M42). Also shown in Tab.~\ref{tab:dyn} are the initial masses for
these stars, $m_\mathrm{ini~new}$, derived by matching the luminosity and 
effective temperature of the newly calibrated O star spectral types by
\citet{MSH05} with the values from the \citet{MM03} rotating stellar
evolution models for massive stars \citep[for details see][]{W09b}.
Generally, the new spectroscopic mass estimates
from the new calibration agree much better with the dynamical masses
than the old spectroscopic mass estimates. For the analysis done in
this work for WR20a A and $\Theta$ Orionis C1 the dynamical
masses are used for the old calibration and $m_\mathrm{ini~new}$ for
the new one.

\begin{table*}
\centering
\caption{\label{tab:dyn} For these massive stars in star clusters
  dynamical mass estimates exist from the orbits of binaries.}
\begin{tabular}{cccccccc}
Cluster&Star&Sp Type&$m_{\rm dynamical}$&$m_{\rm old}$&$m_{\rm
  new}$&$m_\mathrm{ini~new}$&Ref.\\
&&&$M_\odot$&$M_\odot$&$M_\odot$&$M_\odot$&\\
\hline
Trumpler 14/16&FO15B&O9.5V&16.0 $\pm$ 1.0& 23.3 $\pm$ 2.0&16.5 $\pm$ 1.5& 17.9 -3.9/+3.1& (1)\\
Trumpler 14/16&FO15A&O5.5V&30.0 $\pm$ 1.0& 50.4 $\pm$ 6.0&34.2 $\pm$ 3.0& 37.7 -3.7/+7.3& (1)\\
M42&$\Theta$ Orionis C1&O6Vpe&35.8 $\pm$ 7.2&45.0 $\pm$ 5.0&31.7 $\pm$
6.0&34.3 -4.3/+4.7&(2)\\
Trumpler 14/16&HD93205A&O3V&56.0 $\pm$ 4.0&87.6 $\pm$ 12.0&58.3 $\pm$ 10.0& 64.6 -4.6/+5.4&(3)\\
Westerlund 2&WR20a A&WN6ha&82.7 $\pm$ 5.5&-&-&121.0 -41.0/+29.0&(4)\\
NGC3603&NGC3603-A1&WN6ha&116.0 $\pm 31$&120.0 $\pm$ 15.0&-&121.0 -41.0/+29.0&(5)\\
\end{tabular}

$^{}1$ \citet{NMF06},
$^{2}$ \citet{KWB08},
$^{3}$ \citet{MBN01},
$^{4}$ \citet{NRM08},
$^{5}$ \citet{SMS08}
\end{table*}

\subsection{The cluster sample}
\label{sub:old}

\begin{figure}
\begin{center}
\includegraphics[width=8cm]{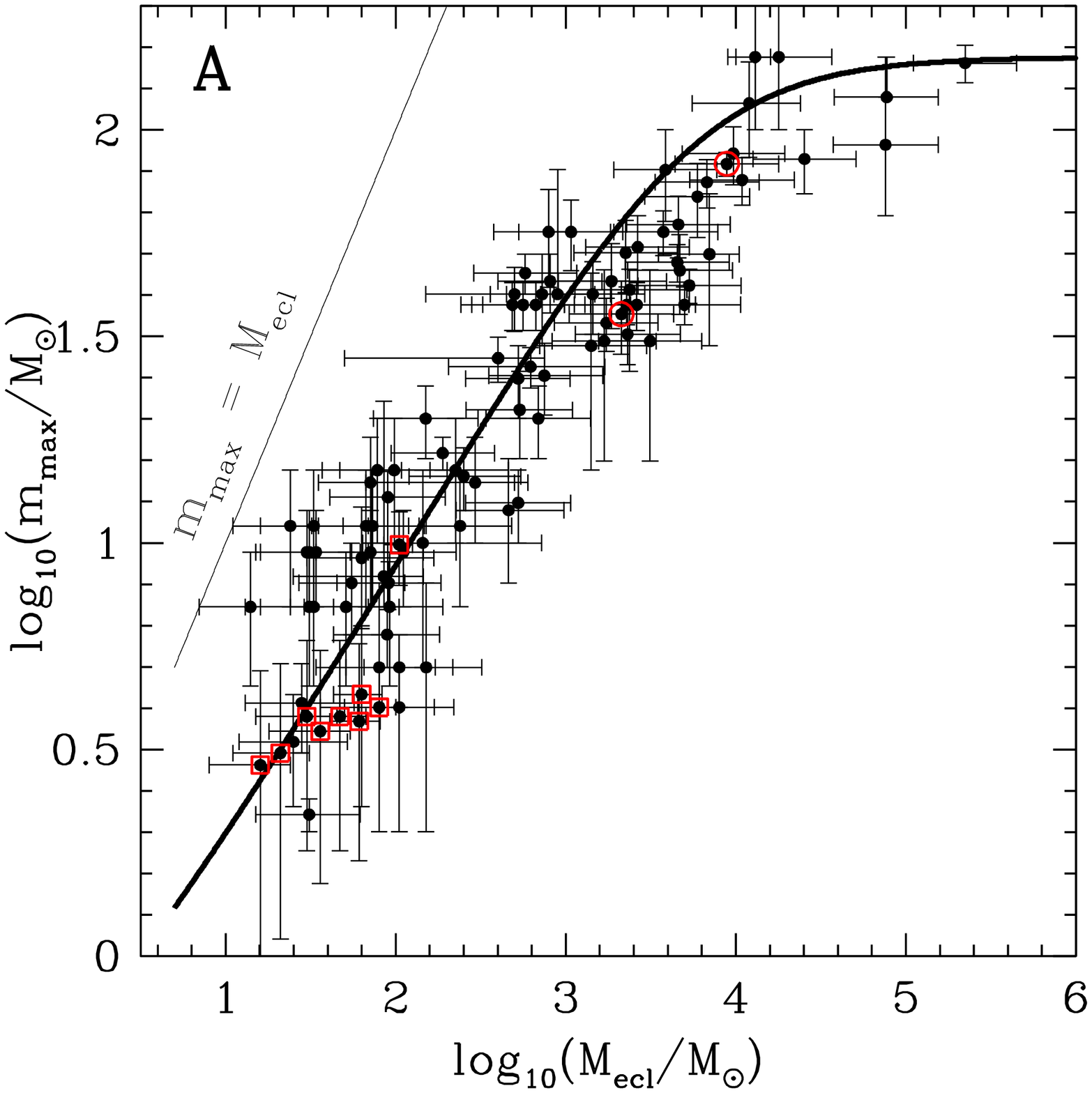}
\includegraphics[width=8cm]{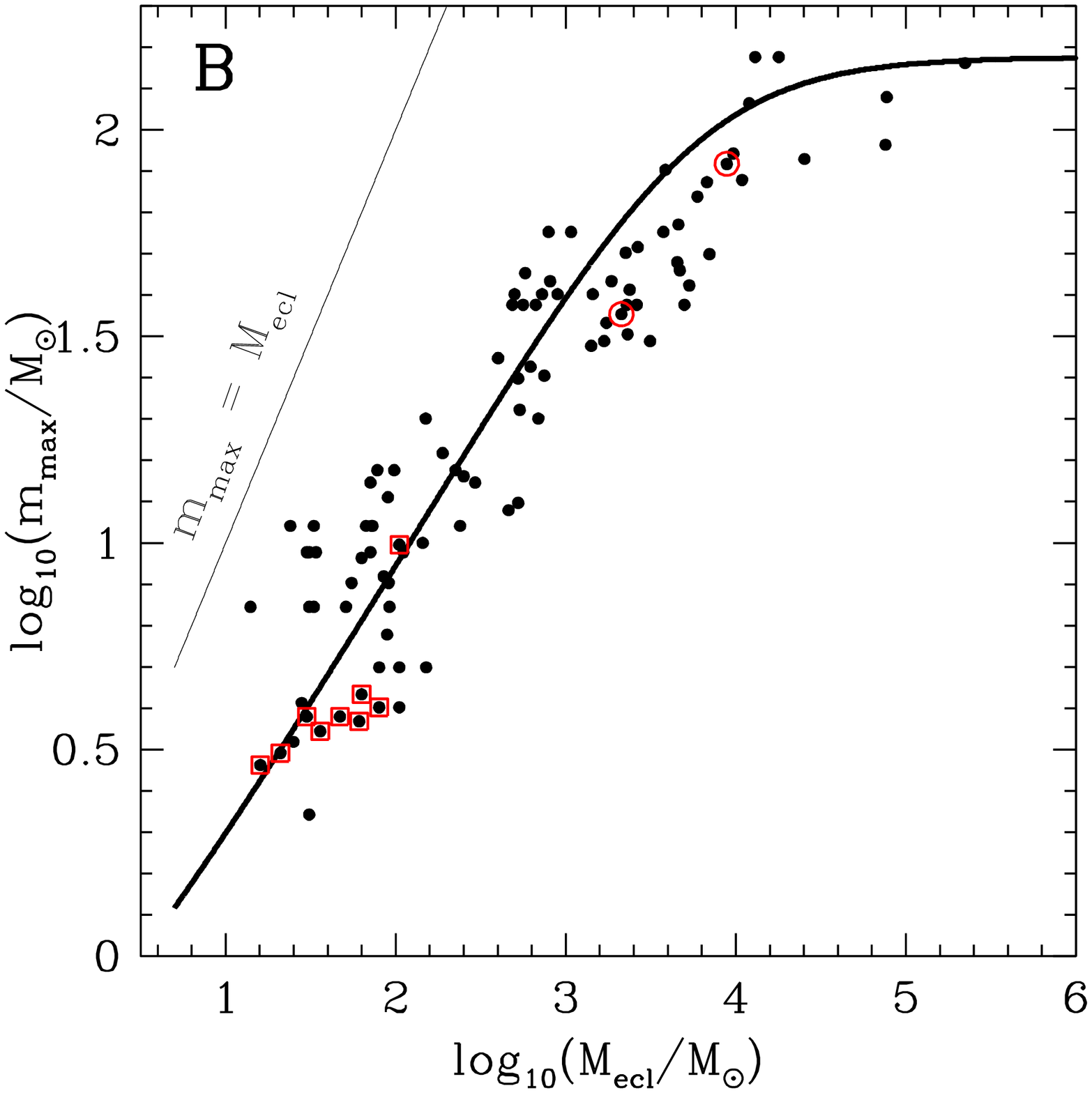}
\vspace*{-2.0cm}
\caption{Plot of the most-massive-star vs. star-cluster-mass data using the
  literature values for the stellar masses and the old effective
  temperature scale calibration for O-stars. {\it Panel A}: With error
  bars and {\it panel B} without error bars. The {\it thick
  solid line} represents the analytic model from \citet{WK04}, see
  also Fig.~\ref{fig:mvM_hist}.
  The literature values with {\it circles} around them
  have dynamical mass estimates for the most-massive star while the
  ones with {\it boxes} are the sample of \citet{FMT09} for
  which the masses of the most-massive stars are only indirectly
  calculated. The {\it thin solid line} marks the identity when a
  cluster is made of only one star.}
\label{fig:MvM1}
\end{center}
\end{figure}

Table~\ref{tab:clustersold} in Appendex~\ref{app:data} includes the
\citet{WK05b} sample of most-massive stars in star clusters together
with the new entries compiled here from the literature. Fig.~\ref{fig:MvM1}  
shows the most-massive-star vs.~star-cluster-mass relation from this
table using the old stellar masses for the O stars. Futhermore, the
Fig.~shows the theoretical analytic result (the {\it thick solid
line}) from \citet{WK04}, which numerically solves eqs.~\ref{eq:norm}
and \ref{eq:Mecl} but with the canonical multi-part power-law IMF and
assuming a fundamental upper mass limit for stars, $m_\mathrm{max *}$
= 150 $M_\odot$ to arrive at a relation for $m_\mathrm{max} =
f_\mathrm{ana}(M_\mathrm{ecl})$.

\begin{figure}
\begin{center}
\includegraphics[width=8cm]{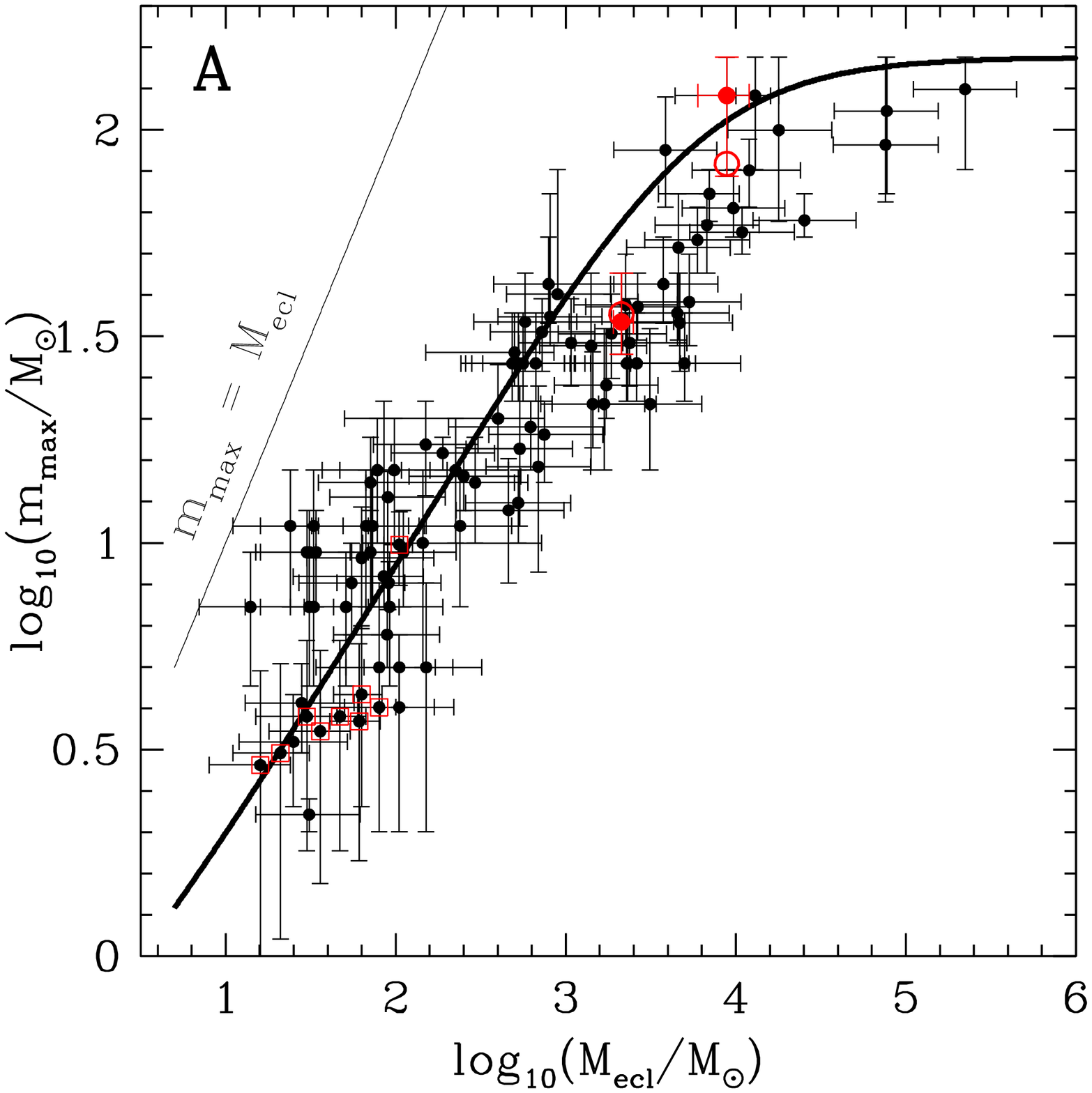}
\includegraphics[width=8cm]{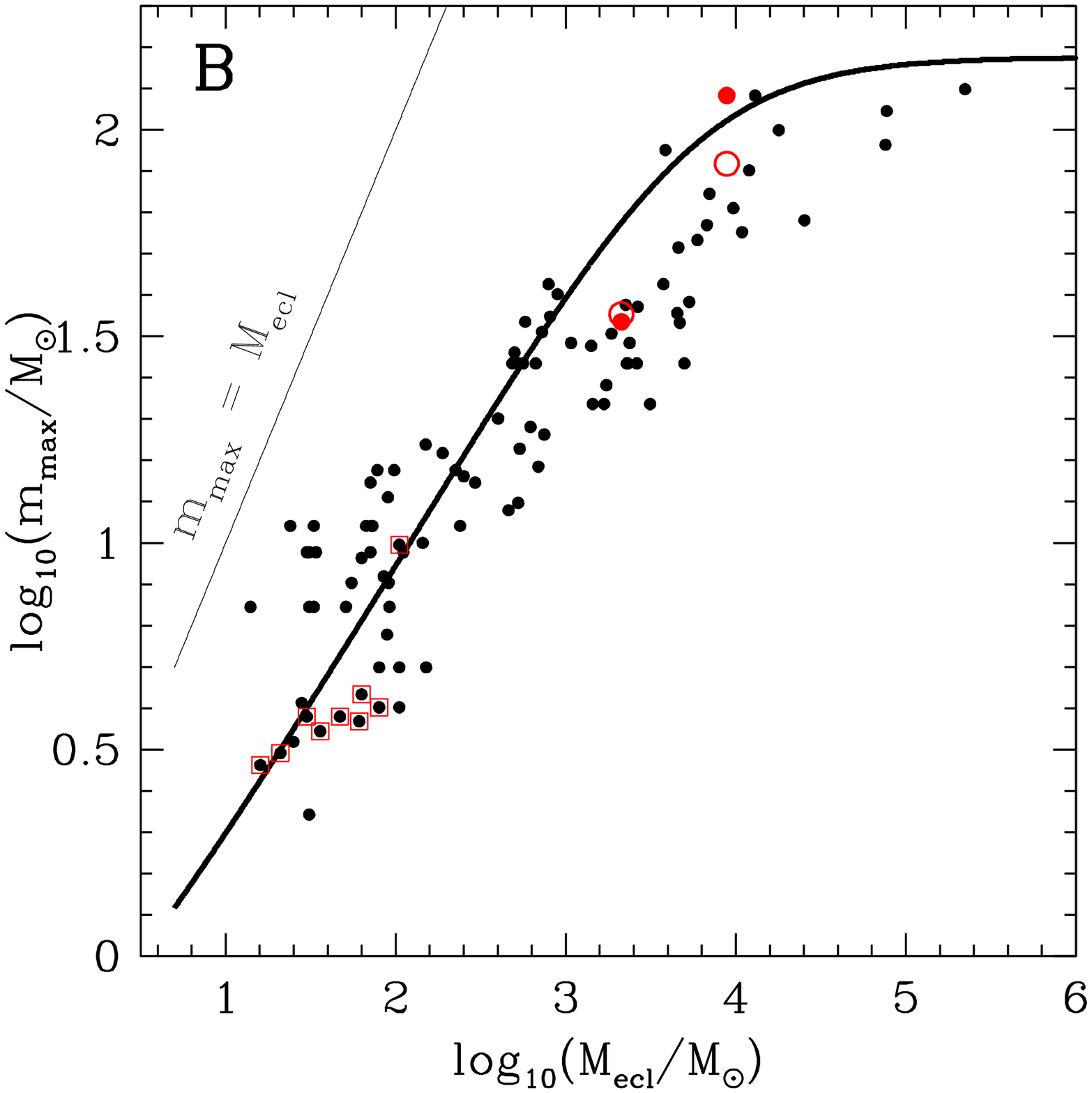}
\vspace*{-2.0cm}
\caption{Like Fig.~\ref{fig:MvM1} but using the new effective
  temperature-scale calibration for O-stars from \citet{MSH05}
  corrected for evolutionary effects to initial masses
    \citep{W09b}. {\it Panel A}: With error bars and {\it panel B}
  without error bars. The dynamical mass estimates are shown only
  as {\it open circles} as the evolutionary correted initial masses
  are used for the actual analyses.}
\label{fig:MvM2}
\end{center}
\end{figure}

The masses of the most-massive stars derived from the new spectral
type to mass conversion are shown in Fig.~\ref{fig:MvM2}
together with the same lines as in Fig.~\ref{fig:MvM1}. 

\subsubsection{Errors}
The error bars for $M_\mathrm{ecl}$ in Tab.~\ref{tab:clustersold} are
either directly taken from the respective literature source, or by
assuming an uncertainty of 100\% in the number
of observed stars. For $m_\mathrm{max}$ the errors are
again taken either directly from the literature or the spectral type
is converted to mass by the \citet{VGS96} tables for column \# 3 and
\citet{MSH05} for column \# 4\footnote{The masses are corrected for
  stellar evolutionary effects as described in \citet{W09b}.}
and the spectral subtype +1 and -1 is
used as the upper and lower limit for the stellar mass\footnote{For
example, for an O4 V star the spectral types O5 V and O3 V are used
to determine the lower and upper mass limit.}, respectively. The
errors in the distance and age are from the literature only. 

\section{Statistical Analyses}
\label{se:stat}

In the supplement of \citet{PAK08} the probability for the {\it i}th
massive star randomly chosen from a number of stars $N$ is
given, assuming $m_\mathrm{max *}$ = 150 $M_\odot$. For
{\it i} = 1 (the most-massive star) the probability is, 

\begin{equation}
\label{eq:pn}
p_{N (m)} = N \left(\int_{m_\mathrm{min}}^{m} \widetilde{\xi}(m) dm \right)^{N-1} \widetilde{\xi}(m),
\end{equation}
with $\widetilde{\xi}(m)~\propto~\xi(m)$ being the probability density
distribution and $\xi(m)$ the IMF as described in Appendix~\ref{app:IMF}.

In order to get the number of stars, $N$, required for eq.~\ref{eq:pn}
for a given cluster with $M_\mathrm{ecl}$, an array of cluster masses
between 5 $M_\odot$ and $10^6$ $M_\odot$ is divided by the mean mass,
$m_\mathrm{mean}$, of the IMF. For the IMF used here (see
appendix~\ref{app:IMF} for details) $m_\mathrm{mean}$ = 0.36 $M_\odot$, if
$m_\mathrm{min}$ = 0.01 $M_\odot$ and $m_\mathrm{max *}$ = 150
$M_\odot$.

\begin{figure}
\begin{center}
\includegraphics[width=8cm]{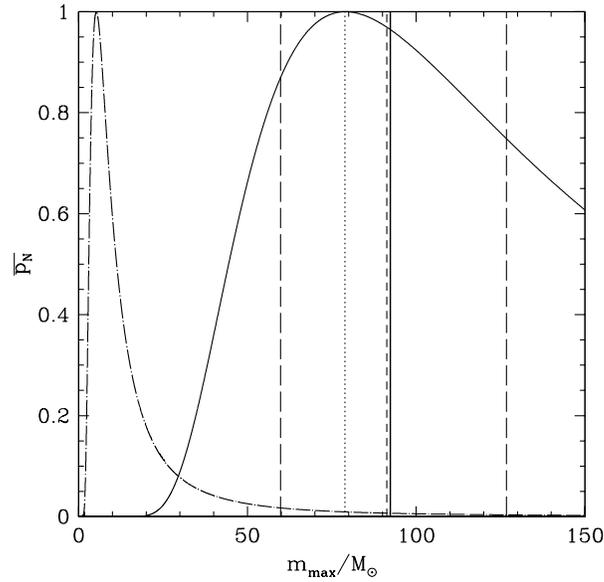}
\vspace*{-2.0cm}
\caption{The {\it solid} line shows the distribution of the
most-massive star for $N$ = 11111 ($M_\mathrm{ecl}$ $\approx$ 4000 $M_\odot$) 
and the {\it dash-dotted} line is the same for $N$ = 278 ($\approx$ 100
$M_\odot$) according to eq.~\ref{eq:pn}. In this plot $p_{N}$ is normalised 
to 1 at the most common (mode) value in order to give both 
curves the same height. For the $N$ = 11111 case the following
statistical properties are also shown: the {\it vertical dotted line}
is the mode, the {\it vertical dashed line} is the median, the {\it
vertical solid line} is the expectation value and the two {\it vertical
long-dashed lines} mark the 1/6th and 5/6th quantiles. See text for further
details.} 
\label{fig:rand_dist}
\end{center}
\end{figure}

Fig.~\ref{fig:rand_dist} shows the distribution obtained by random
sampling (eq.~\ref{eq:pn}) for two examples of $N$ = 278 ({\it
dash-dotted line}) and $N$ = 11111 ({\it solid line}). For each $N$
the five statistical values are calculated. 
\begin{itemize}
  \item The arithmetic mean or expectation value, marked as a {\it
    solid vertical line} for the $N$ = 11111 case in
    Fig.~\ref{fig:rand_dist}, is the sum of all most-massive stars
    divided by the number of clusters. 
  \item The mode value ({\it dotted vertical line} in
    Fig.~\ref{fig:rand_dist}) marks the most common value (the peak in
    Fig.~\ref{fig:rand_dist}) of the distribution.
  \item The median value ({\it dashed vertical line} in
    Fig.~\ref{fig:rand_dist}) is the value which divides the
    distribution in two. 50\% of the values are above the median while
    50\% are below.
  \item The 1/6th quantile (the left {\it long-dashed vertical line} in
    Fig.~\ref{fig:rand_dist}) is the value below which 1/6th of data
    points lie.
  \item The 5/6th quantile (the right {\it long-dashed vertical line} in
    Fig.~\ref{fig:rand_dist}) is the value above which 5/6th of data
    points lie.
\end{itemize}

The 1/6th and 5/6th quantiles define the region within which lie two
thirds of the most-massive stars lie for random sampling of stars from
the IMF (eq.~\ref{eq:pn}).

\subsection{Completeness of the sample}
\label{sub:tests}
The completeness of the cluster sample presented here strongly depends
on the total number of star clusters expected for the Milky Way (MW)
which are younger than 4 Myr. This depends on the assumed current
star-formation rate (SFR) of the MW \citep[0.8 - 13 $M_\odot$
yr$^{-1}$,][and references therein]{DHK06}, the slope of the embedded  
cluster mass function \citep[$\beta$ = 1.8 - 2.3,][]{LL03}, where
$dN_\mathrm{ecl} = M_\mathrm{ecl}^{-\beta}dM_\mathrm{ecl}$ is the
number of just formed embedded clusters with stellar mass in the
interval $M_\mathrm{ecl}$, $M_\mathrm{ecl} + dM_\mathrm{ecl}$, and the
assumed lower mass limit for star clusters \citep[5 to 100
$M_\odot$,][]{WK05b}. With the observationally favoured parameters
being $SFR$ = 4.0 $M_\odot$ yr$^{-1}$, $\beta$ = 2.0 and
$M_\mathrm{ecl~min}$ = $5 M_\odot$. The total number of young star
clusters in the MW lies therefore between $10^4$ and $10^6$
clusters. The majority of these have masses less than 100 $M_\odot$
and any surveys of them are severely incomplete. For a completeness estimate
we therefore restrict ourselves to clusters more massive than 1000
$M_\odot$ as they are far fewer in numbers and more easily identified
in the MW. For the whole range of parameters of the
MW the number of young star clusters more massive than 1000 $M_\odot$
lies somewhere between 160 and 4452, with 1478 being the value for the
observationally favoured parameters. The sample shown in  
Tab.~\ref{tab:clustersold} includes 30 (-5/+6) clusters which are in
the MW and more  massive than 1000 $M_\odot$ within the uncertainties. This
suggests that between 18.8\% (-3.1/+3.7) and 0.7\% ($\pm$ 0.1) of all such
clusters are in the sample, with 2.0\% (-0.3/+0.4) for the favoured
parameters. Therefore, one has to keep in mind that any statistical
results are possibly limited by the incompleteness of the cluster sample.

\subsection{Statistical tests}
\begin{figure}
\begin{center}
\includegraphics[width=8cm]{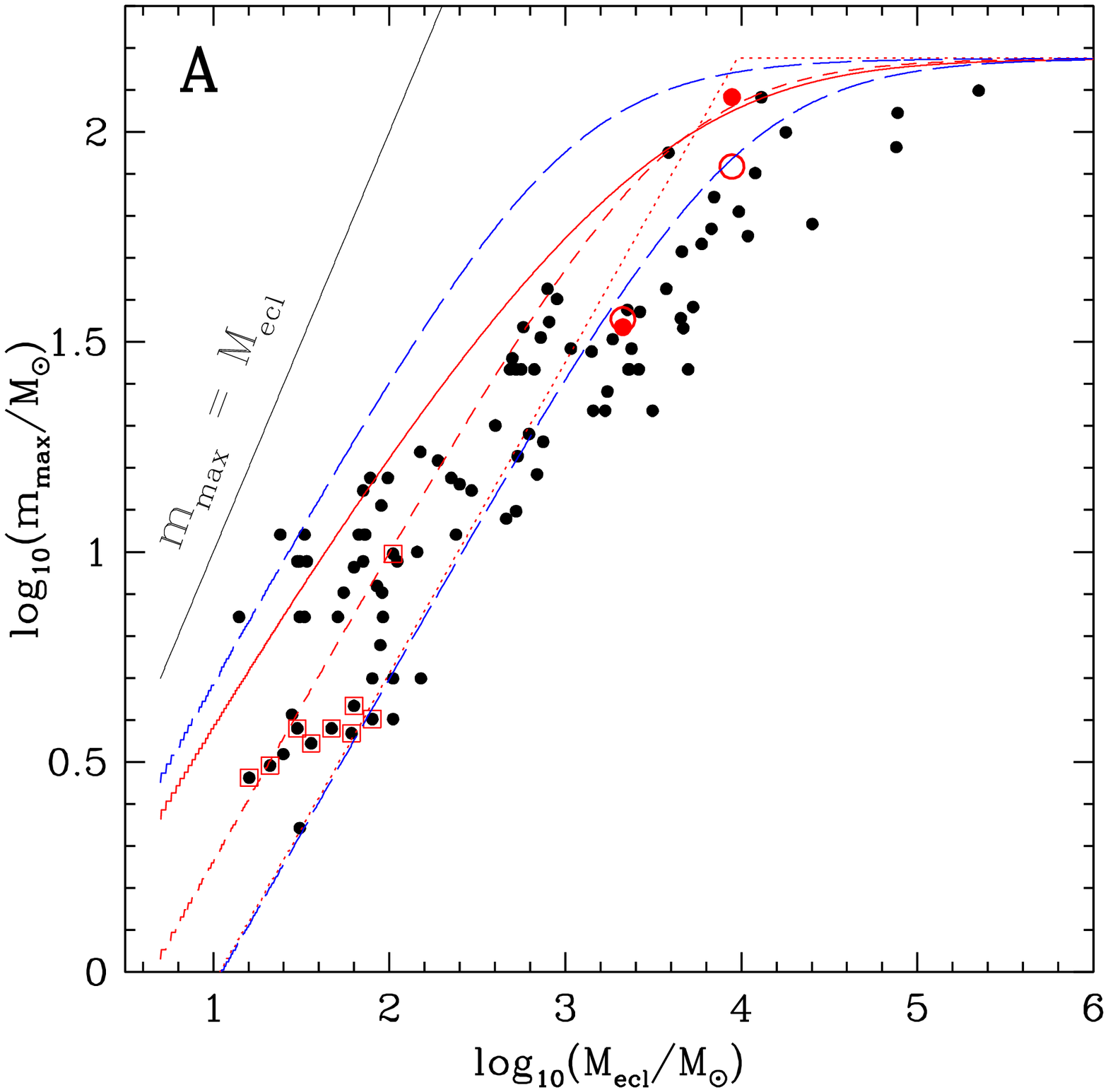}
\includegraphics[width=8cm]{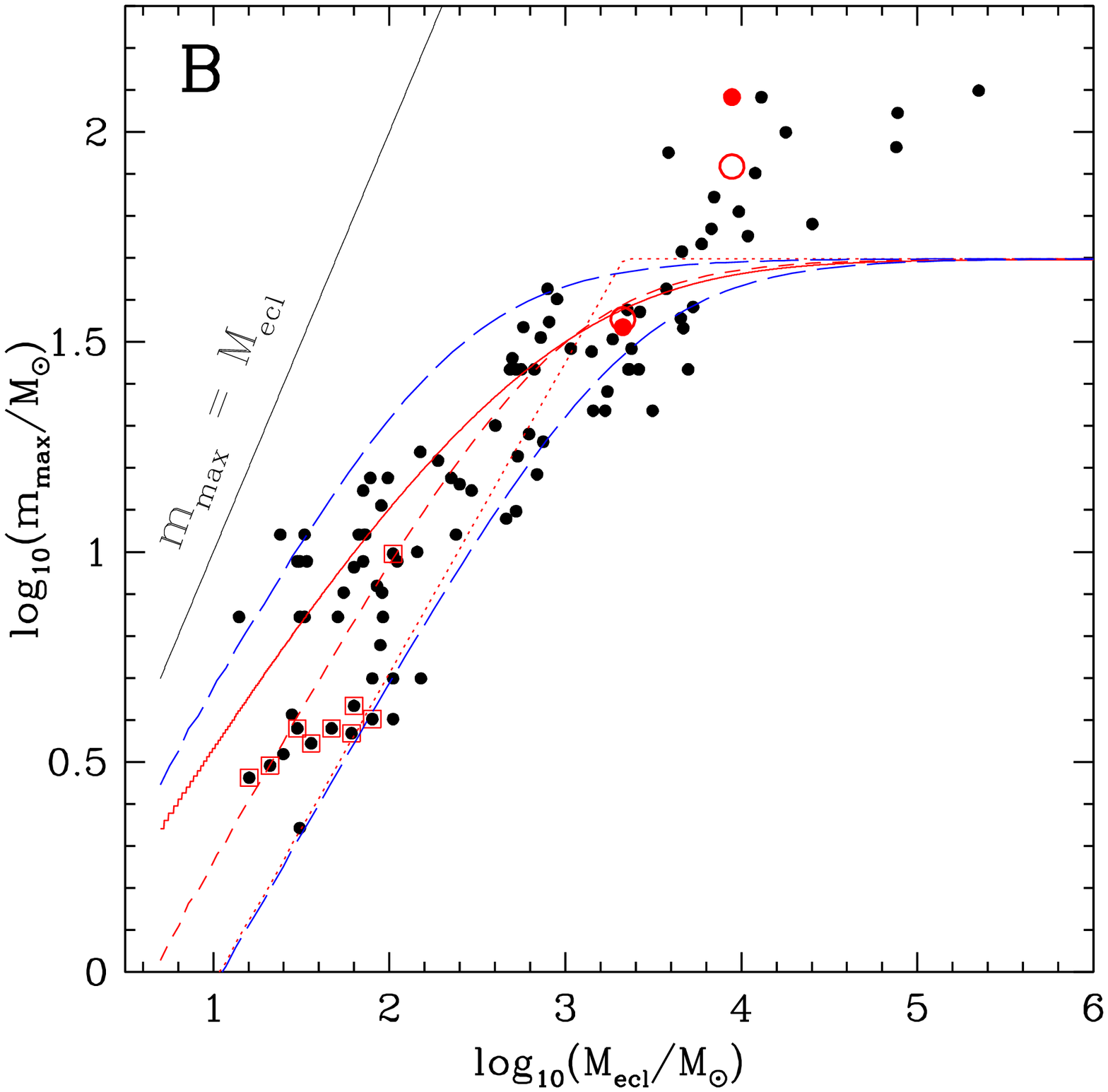}
\vspace*{-2.0cm}
\caption{{\it Panel A}: Most-massive star vs cluster mass. The dots
  are the observed values from column \# 4 from
  Tab.~\ref{tab:clustersold}. The two {\it open circles} indicate
  existing dynamical estimates for the present-day mass of the
  most-massive stars. The {\it boxed} data are from the sample of
  \citet{FMT09}. The {\it dotted} line refers to the mode value for
  random sampling, the {\it short-dashed} line to the median value, the
  curved {\it solid} line marks the mean value and the two {\it
    long-dashed} lines are the 1/6th (lower) and 5/6th (upper) quantiles
  between which 2/3rd of the data points should lie if they were
  randomly sampled from the IMF over the mass range 0.01 $M_\odot$ to
  $m_\mathrm{max *}$ = 150 $M_\odot$. The thin {\it
    solid} line to the left marks the identity where the cluster is
  made-up only of one star. {\it Panel B}: The same as {\it
panel A} but assuming a fundamental upper mass limit for stars of
  $m_\mathrm{max *}$ = 50 $M_\odot$ instead of 150 $M_\odot$.}
\label{fig:stat}
\end{center}
\end{figure}

In {\it panel A} of Fig.~\ref{fig:stat} the mode, mean, median and 1/6th
and 5/6th quantiles for a fundamental upper mass limit of
$m_\mathrm{max *}$ = 150 $M_\odot$ are shown together with the data
points from column \# 4 and the clusters from column \# 3 from
Tab.~\ref{tab:clustersold} which have not been changed by the 
re-calibration. 

Three different statistical tests are applied to the data in order to
verify whether or not the observed most-massive stars are consistent
with being randomly drawn from the IMF.

\subsubsection{Percentage of stars between the 1/6th and 5/6th quantiles}

\begin{figure}
\begin{center}
\includegraphics[width=8cm]{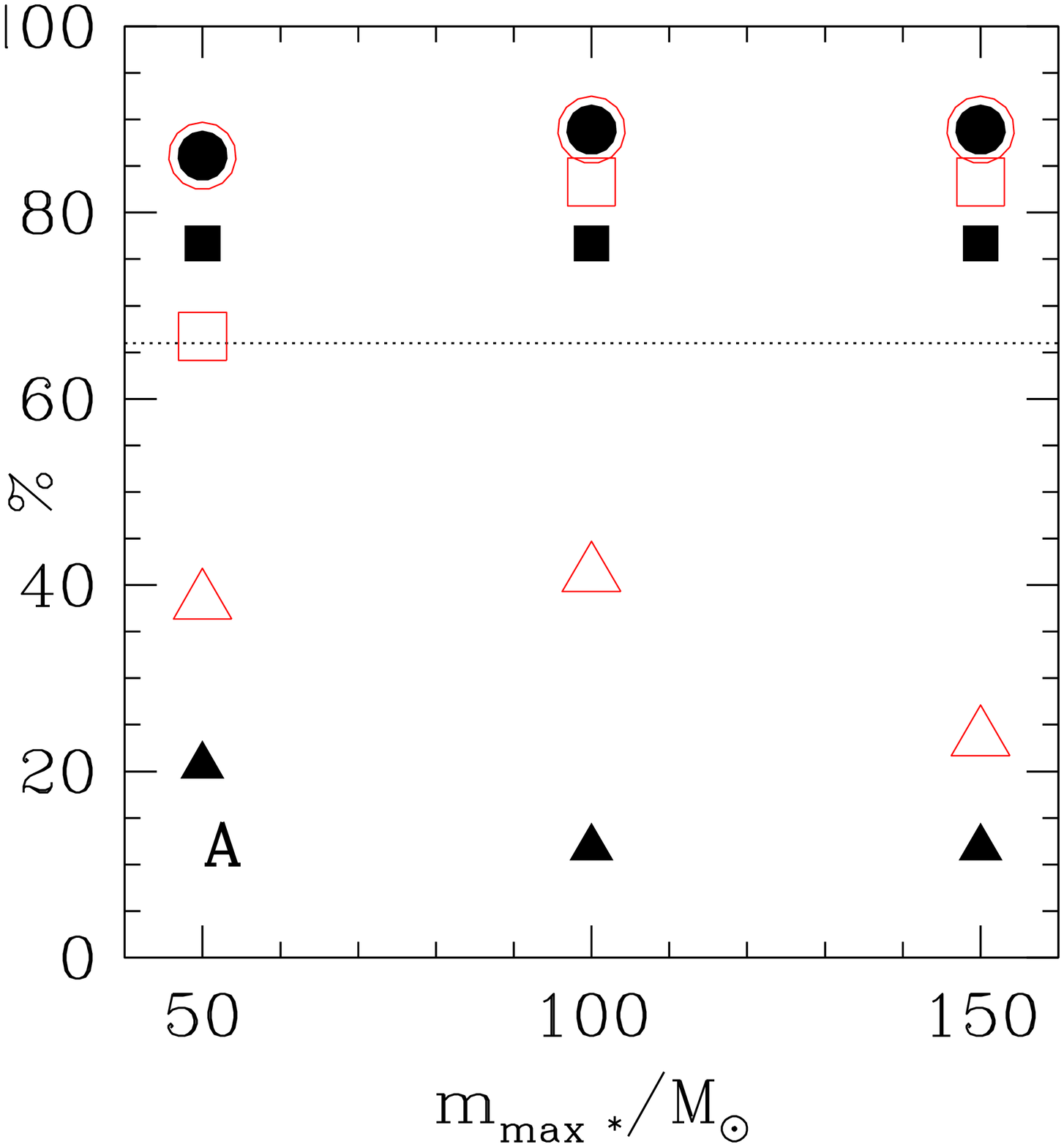}
\includegraphics[width=8cm]{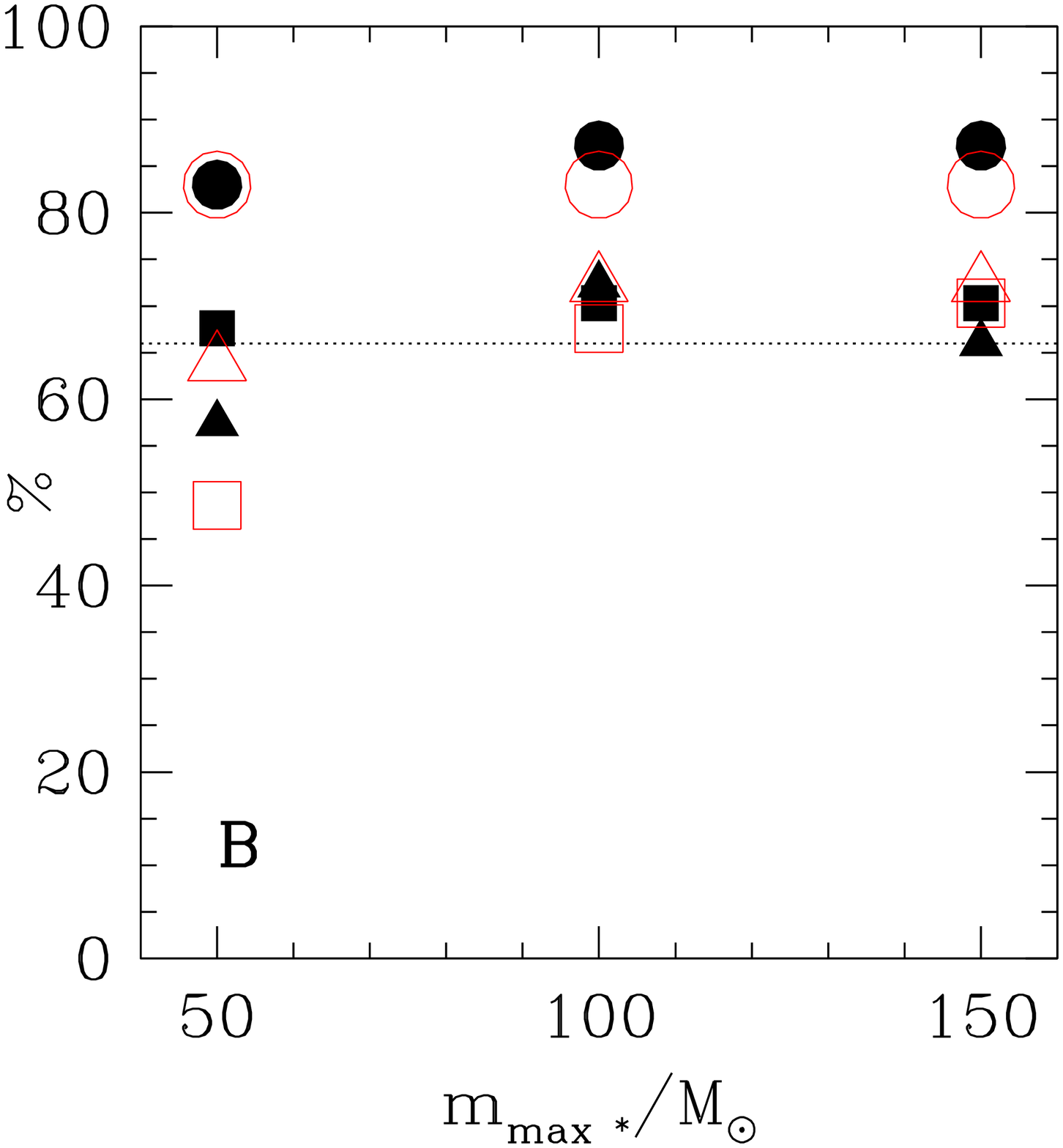}
\vspace*{-1.5cm}
\caption{Percentage of most-massive stars within the 1/6th to 5/6th
  quantiles for three different assumptions of the fundamental upper
  mass limit, $m_\mathrm{max *}$, of 50, 100 and 150 $M_\odot$. {\it
    Solid symbols} refer to the new O star calibration in
  Tab.~\ref{tab:clustersold} and {\it open symbols} to the old
  one. {\it Circles} mark the percentages for clusters below 
  $10^2 M_\odot$ while {\it squares} are for 100 $< M_\mathrm{ecl}
  \le 10^3 M_\odot$ and {\it triangles} mark
  clusters more massive than $10^3 M_\odot$. {\it Panel A}: These
  values do not take into account the errors in $M_\mathrm{ecl}$ and
  $m_\mathrm{max}$ in Tab.~\ref{tab:clustersold} while in {\it panel
    B} these errors are included. The horizontal {\it dotted line} at
  66\% marks the fraction of stars that ought to lie between the 1/6th
  and 5/6th quantiles if random sampling between 0.01 $M_\odot$ and
  $m_\mathrm{max *}$ were true.}
\label{fig:stat2}
\end{center}
\end{figure}

As is visible in this Fig.~\ref{fig:stat} there is a general
change in behaviour of the data points around a cluster mass of about
100 $M_\odot$ and around 1000 $M_\odot$ with respect to what is
expected from random sampling. Below the 100 $M_\odot$ limit the data
show a larger spread while above 1000 $M_\odot$ the slope of the
$m_\mathrm{max}$--$M_\mathrm{ecl}$-relation changes. {\it
Panel A} of Fig.~\ref{fig:stat2} shows the percentage of
the most-massive stars within the 1/6th and 5/6th quantiles in three
samples, one for the clusters below 100 $M_\odot$, one for the
clusters between 100 and 1000 $M_\odot$ and one for the
ones above 1000 $M_\odot$. Additionally, the figure shows the same
numbers for different assumptions on the fundamental upper mass limit
($m_\mathrm{max *}$) for stars. Here the clusters above 1000
$M_\odot$ (filled and open triangles) are far below the 2/3rd range
which would be expected from random sampling. The clusters below 100
$M_\odot$ (filled and open circles) and the intermediate clusters (100
to 1000 $M_\odot$, filled and open squares) are very tightly within
the 1/6th and 5/6th quantiles. About 90\% and 78\% of the clusters are
within the range, respectively. In {\it panel B} of
Fig.~\ref{fig:stat2} the same is shown but including the error bars
for $m_\mathrm{max}$ and $M_\mathrm{ecl}$ from
Tab.~\ref{tab:clustersold} by making the same calculations as before
but using the minimal and maximal values for $m_\mathrm{max}$ and
$M_\mathrm{ecl}$. The low-mass clusters are still more tightly
distributed within the 1/6th and the 5/6th quantiles than expected.
The intermediate and high-mass cluster seem to be consistent with
random sampling when the maximum effect of the errors is applied to
the data.

\subsubsection{Distribution around the Median}

\begin{figure}
\begin{center}
\includegraphics[width=8cm]{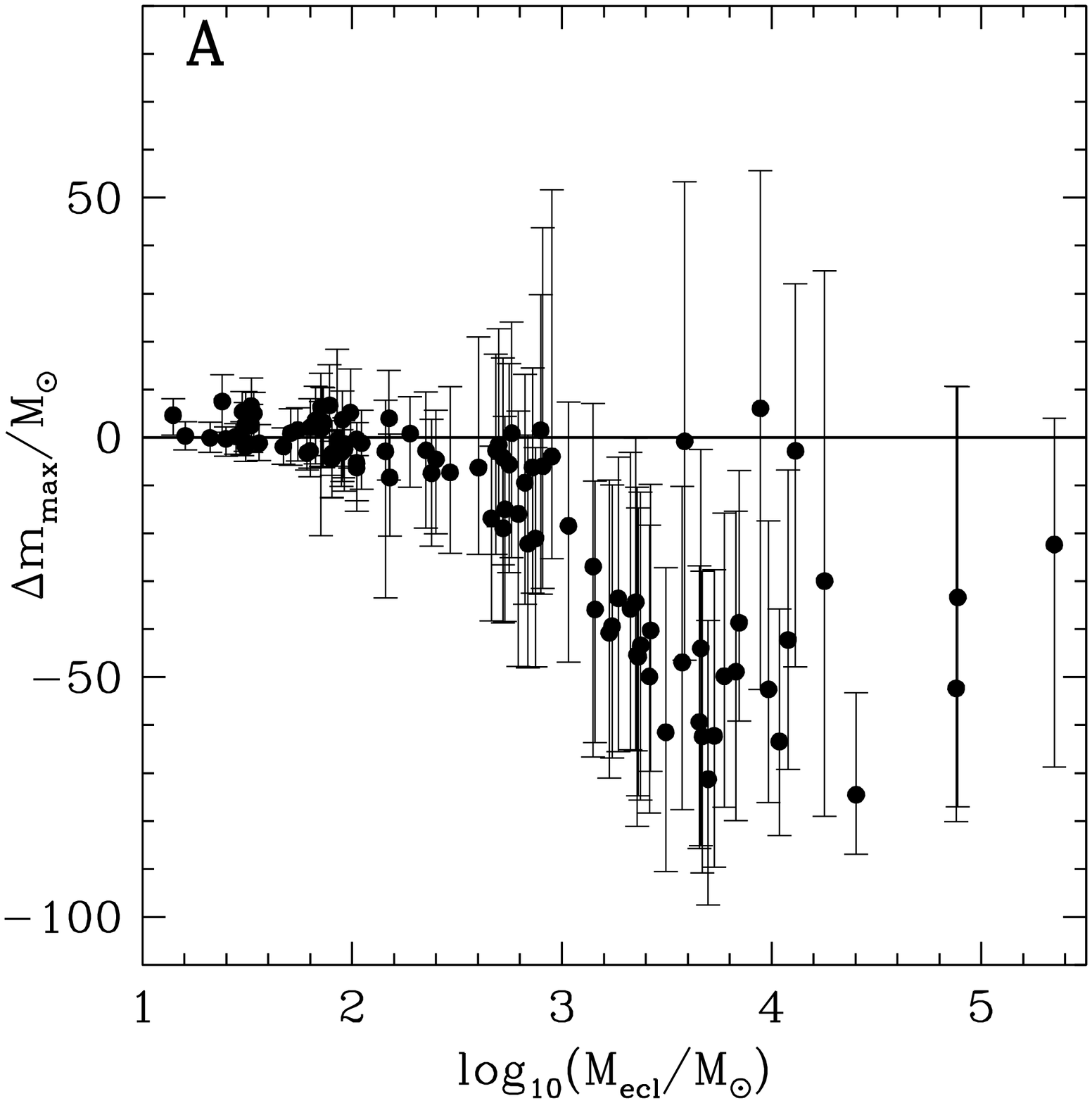}
\includegraphics[width=8cm]{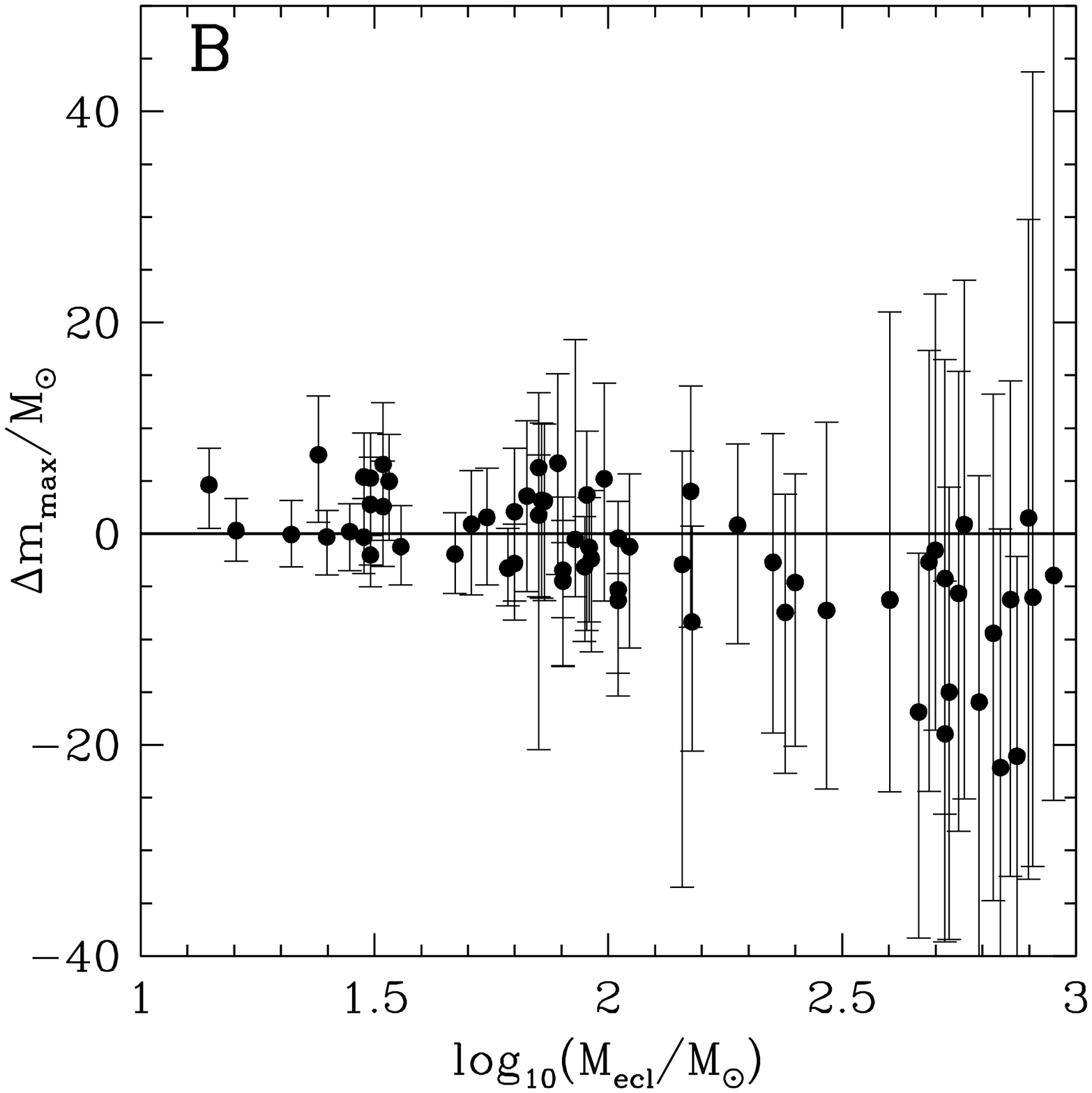}
\vspace*{-2.0cm}
\caption{Distance of the observed most-massive star from the expected
  median for random sampling up to $m_\mathrm{max *}$ = 150
  $M_\odot$. Here the new \citet{MSH05} mass scale is used for O
  stars. {\it Panel A} show the whole sample while {\it panel B}
  contains only the clusters below $10^3 M_\odot$. For random sampling
over the mass range 0.01 $M_\odot$ to 150 $M_\odot$ the data ought to
be distributed symetrically about $\Delta m_\mathrm{max}$~=~0.}
\label{fig:stat3}
\end{center}
\end{figure}

\begin{figure}
\begin{center}
\includegraphics[width=8cm]{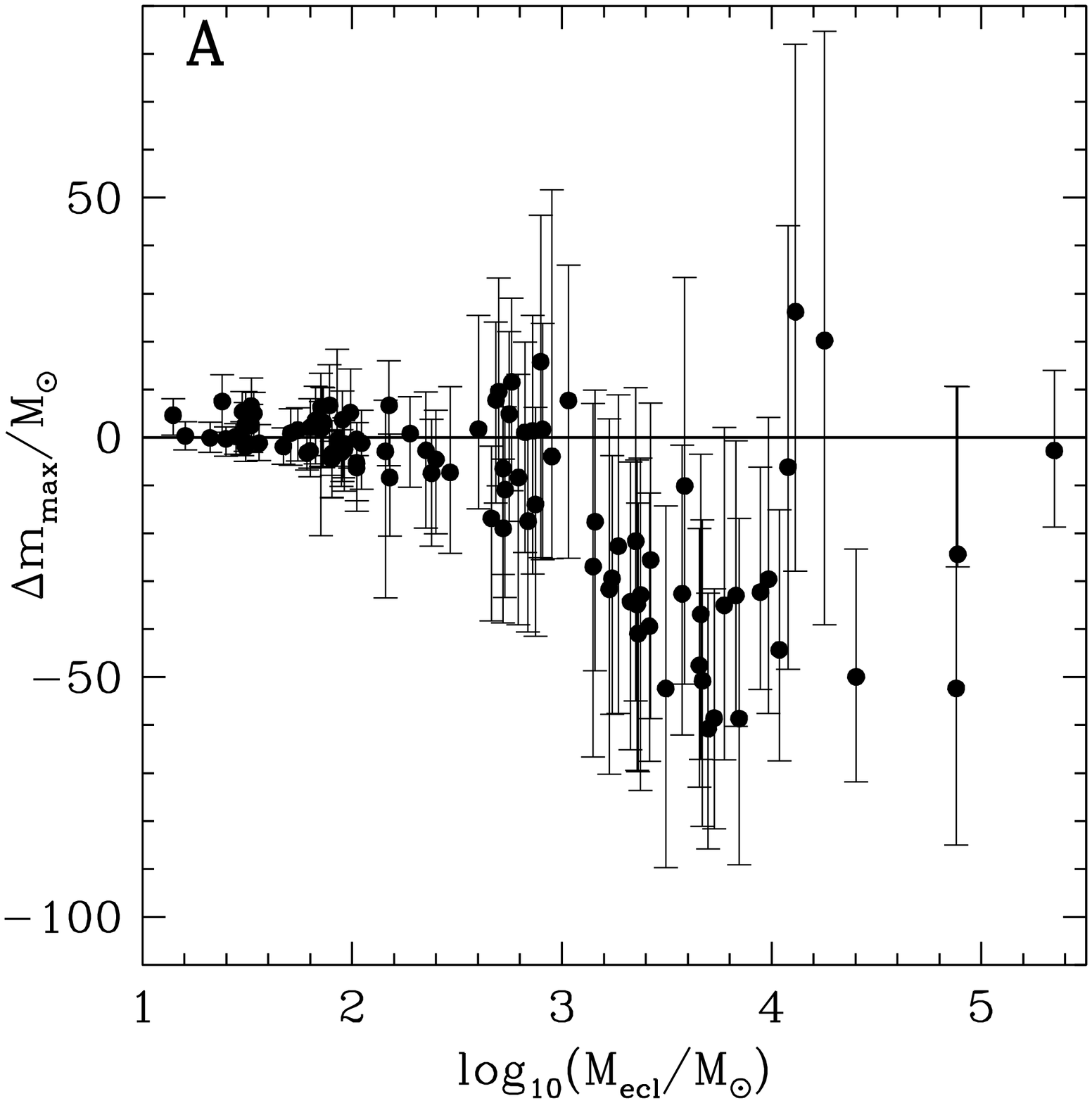}
\includegraphics[width=8cm]{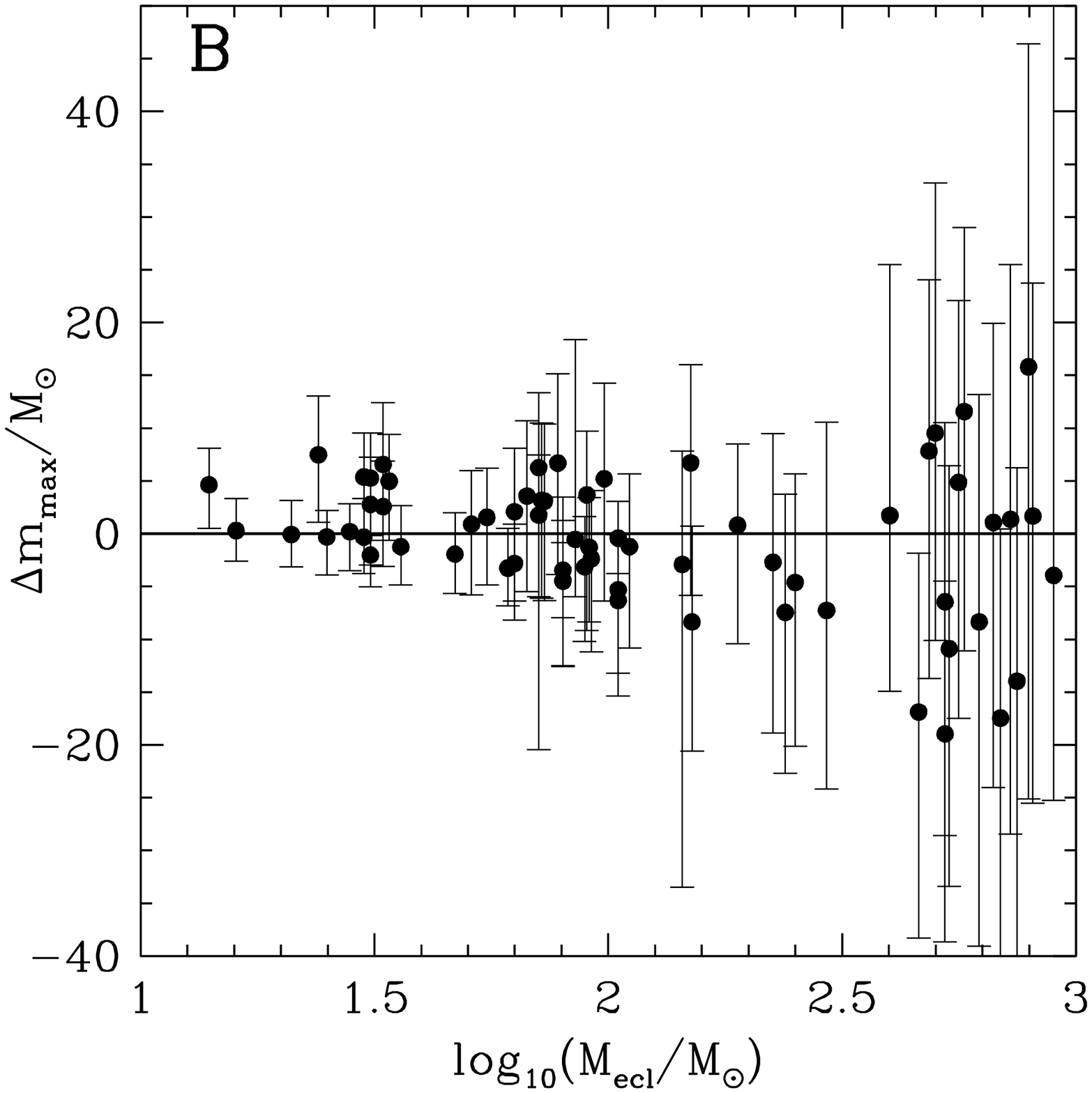}
\vspace*{-2.0cm}
\caption{Like Fig.~\ref{fig:stat3} but for the old O star calibration.}
\label{fig:stat4}
\end{center}
\end{figure}

Also important is the distribution of the $m_\mathrm{max~obs}$ values
around the median of the expected distribution for random
sampling. The median is the statistical value for which 50\% of the
data should lie above and below. For the whole sample 25.7\% are above the
median and 74.3\% below if one uses the new \citet{MSH05} O star mass
scale and assumes a fundamental upper mass limit of $m_\mathrm{max
  *}$ = 150 $M_\odot$. In the sub-sample of clusters below 100 
$M_\odot$ there are 56.8\% above and 43.2\% below the median, for the
clusters with $100 < M_\mathrm{ecl} \le 1000 M_\odot$ there are 13.3\%
above and 86.7\% below the median while for 
the high-mass clusters 2.9\% and 97.1\% are above and below the
median, respectively. In Fig.~\ref{fig:stat3} the distribution of
$m_\mathrm{max~median} - m_\mathrm{max~obs~new}$ is shown for the
whole cluster sample ({\it panel A}) and for the clusters below $10^3
M_\odot$ ({\it panel B}). For the old O star mass scale the
distribution is shown in Fig.~\ref{fig:stat4}. The percentages in the
case of the old O star mass scale are
35.6/64.4\% for the total sample and 56.8/43.2\%, 40.0/60.0\% and
8.8/91.2\% for, respectively, the clusters below 100 $M_\odot$, the
clusters with $100 < M_\mathrm{ecl} \le 1000 M_\odot$ and the clusters
above $10^3 M_\odot$.

\subsubsection{Wilcoxon-Signed-Rank-Test}

The Wilcoxon signed rank test \citep{BJ77}\footnote{A short
  introduction into the test and pre-calculated tables for the
  probabilities for different $N$ can be found at:
  http://comp9.psych.cornell.edu/Darlington/index.htm}, tests
whether or not the data is consistent with being symmetrically
distributed around the median. It reveals for the new calibration a
probability\footnote{$\sim 0.1$ is the highest probability the
  Wilcoxon Signed Rank test allows for.}, $p(M_\mathrm{ecl} \le 100
M_\odot)$, of 0.014 for clusters with masses smaller or equal to
100 $M_\odot$, a $p(100 M_\odot < M_\mathrm{ecl} \le 1000 M_\odot)$ of
$1.9 \cdot 10^{-7}$ for cluster masses between 100 and 1000 $M_\odot$
and a $p(M_\mathrm{ecl} > 1000 M_\odot)$ of $2.8 \cdot 10^{-9}$ for
the clusters above 1000 $M_\odot$. For the old calibrations the
probabilities are $p(M_\mathrm{ecl} \le 100 M_\odot)$ = 0.014, $p(100
M_\odot < M_\mathrm{ecl} \le 1000 M_\odot)$ = 0.035 and
$p(M_\mathrm{ecl} > 1000 M_\odot)$ = $1.2 \cdot 10^{-8}$.

\subsection{Dependence on the high-mass IMF slope}
\label{sub:alt}
 
\begin{figure}
\begin{center}
\includegraphics[width=8cm]{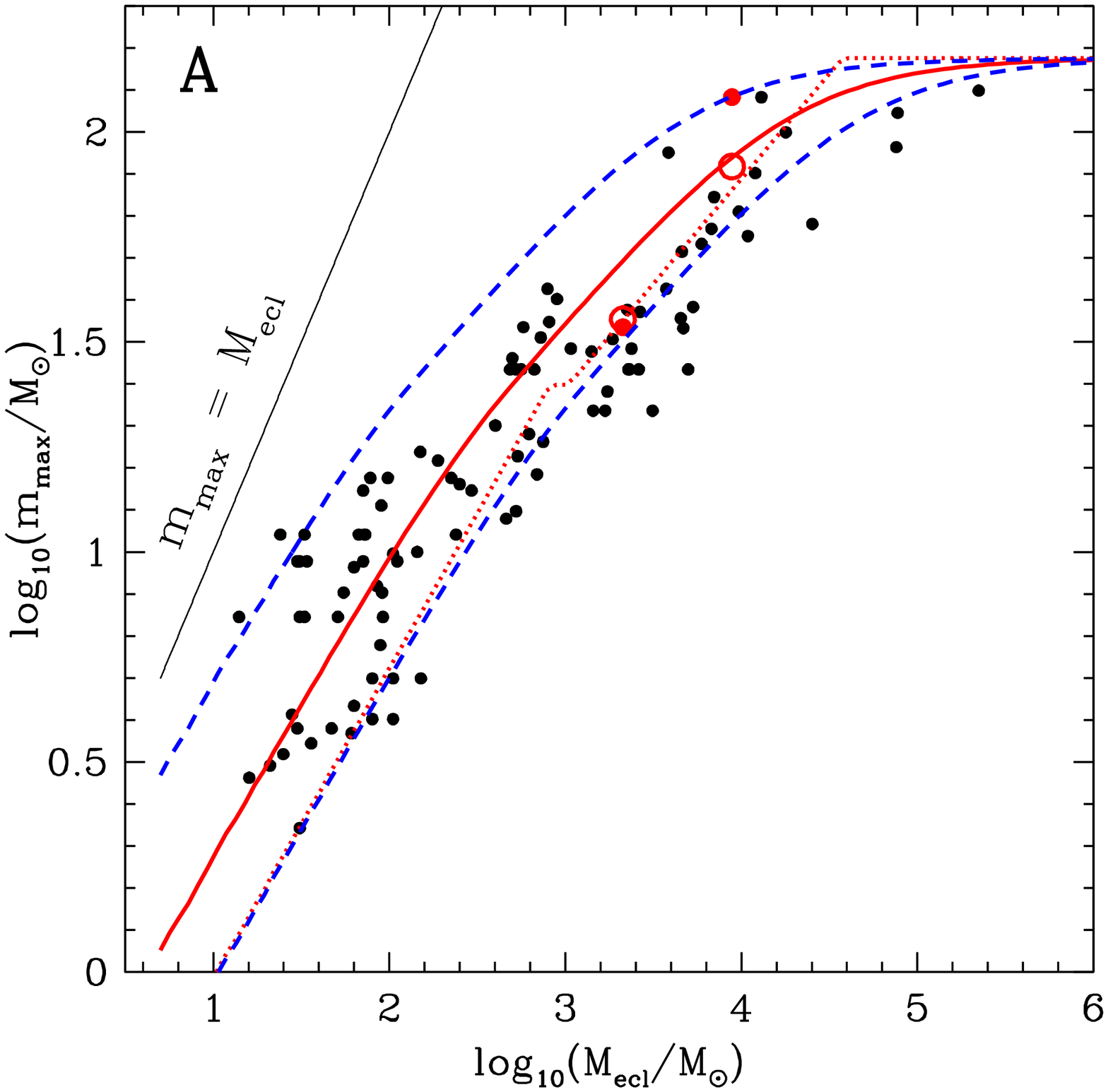}
\includegraphics[width=8cm]{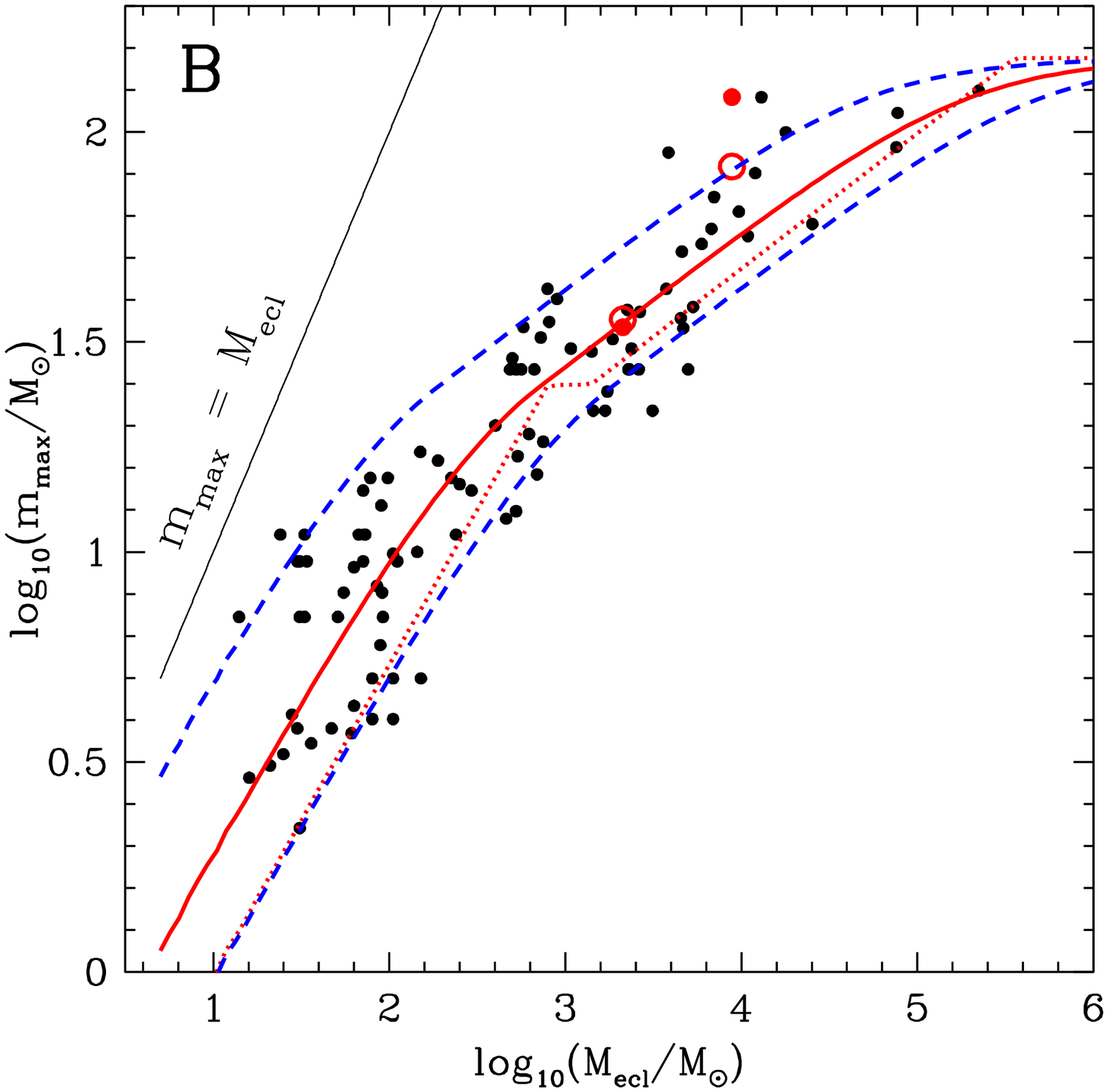}
\vspace*{-2.0cm}
\caption{Like Fig.~\ref{fig:stat} but the IMF slope above 25 $M_\odot$ is
  3.0 ({\it Panel A}) and 4.1 ({\it Panel B}) instead of the Salpeter
  value of 2.35. The {\it solid line} is the median, the {\it dotted
    line} is the mode and the {\it dashed lines} are the 1/6th (lower)
  and the 5/6th (upper) quantiles.}
\label{fig:stat5}
\end{center}
\end{figure}

The general assumption in this paper, that the stars in a
star cluster follow a universal IMF which is characterised by a
Salpeter/Massey slope of 2.35 for all stars above 0.5 $M_\odot$, is
strongly supported by almost all observational evidence (see
appendix \ref{app:IMF} for a list of references). However, if the
IMF-slope for high-mass stars is steeper than 2.35, it is not sure
whether or not a fundemental upper mass limit exists, as was pointed
out by  \citet{OC05}. Fig.~\ref{fig:stat5} shows the mean, median,
mode, 1/6th and 5/6th quantiles for two different assumptions of the
high-mass slope of the IMF. In {\it Panel A} the slope for stars more
massive than 25 $M_\odot$ is changed to $\alpha_3$ = 3.0 and in {\it
  Panel B} to $\alpha_3$ = 4.1, while $\alpha_{2}$ = 2.35 for stars
between 0.5 and 25 $M_\odot$ in both cases. 25 $M_\odot$ was chosen
because it is the $m_\mathrm{max}$ value of the 
plateau like feature for clusters with masses between $10^3$ and $4
\cdot 10^3$ $M_\odot$. Only for a slope as steep as $\alpha_3$ = 4.1
are 50\% of stars above and below the median. A slope steeper than $\alpha_3$ =
3.7 is needed in order to have more than 60\% of the stars within the 1/6th
and 5/6th quantiles. Such steep slopes for the high-mass IMF within
star clusters are clearly ruled out by the current state of
observations \citep{Mass98,Elme99,Kr01,La02b}.

\subsection{Dependence on the environment}
\label{sub:pfalz}
\begin{figure}
\begin{center}
\includegraphics[width=8cm]{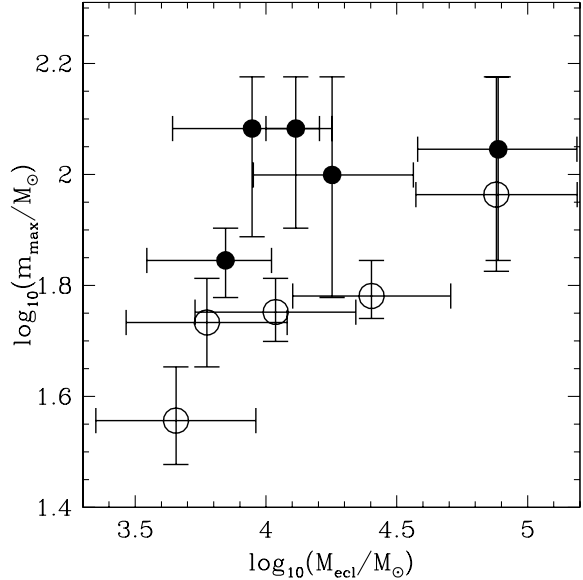}
\vspace*{-2.0cm}
\caption{Like Fig.~\ref{fig:MvM2} but only the clusters which are in
  common with the \citet{Pa09} sample. {\it Filled circles} are
  clusters with radii smaller than 1 pc while the {\it open circles}
  are the clusters with larger radii.} 
\label{fig:pfalz}
\end{center}
\end{figure}

Very recently, \citet{Pa09} studied the dissolution behaviour of young
(1 to 20 Myr) massive star clusters (2000 to 50000 $M_\odot$). She
found that her sample of 23 clusters can be divided into two groups,
loose clusters ($R_\mathrm{ecl} >$ 1 pc) and tight clusters ($R_\mathrm{ecl}
<$ 1 pc), where $R_\mathrm{ecl}$ is her estimated cluster radius. The
radii of the groups each follow a rather tight sequence 
with time. While the tight clusters expand from $\sim 0.5$~pc to 3~pc
the loose ones evolve from 4 pc to 20 pc on the same time scale,
parallel to the tight ones. Of these 23 clusters 10 are included in
our cluster sample. 5 of them are tight clusters ([OBS 2003] 179,
Westerlund 2, NGC 3603, Trumpler 14, Arches) and 5 are loose ones
(NGC 7380, NGC 2244, IC 1805, NGC 6611, Cyg OB2). When comparing the
most-massive stars against the cluster mass of these two subsets, as
is done in Fig.~\ref{fig:pfalz}, it seems that the clusters which form
tighter and are therefore more dense, have on average a more massive
maximal star, while the loose clusters prefer less massive maximal
stars. For the tight subset a linear function can be fitted with a
slope of 0.09 $\pm$ 0.39 and a rather low linear correlation
coefficient of 0.35 $\pm$ 0.47. The slope for the loose sample
is 0.27 $\pm$ 0.16, somewhat steeper than for the tight sample within
the error bars, 
but the linear correlation coefficient is much larger, about 0.92 $\pm$ 0.08 . 
The combined sample has a slope of 0.22 $\pm$ 0.23 with a linear
correlation coefficient of 0.52 $\pm$ 0.47. Therefore, the difference
in slopes might be indicating a physical dependence of the mass of the
most massive star not only on the cluster mass (previous sections) but
also on the cluster density. But the large error bars make a more
definite statement difficult. Also it should be noted here that the
$R_\mathrm{ecl}$ estimates for the all the loose clusters of the
\citet{Pa09}-sample are the measured median distances of early B type
stars in theses clusters \citep{WSD07} and therefore might not be
directly comparable to radii arrived at with different methods.

Low-mass young clusters are found to be generally small
\citep[$\simless\,1$~pc,][]{TPN98,GMP05,TLK05,RJS06,SHG07}
so it is unclear if and how such a correlation between the most-massive
star and the cluster radius extends to lower masses.

\subsection{A simple Model}
\label{sub:mod}

A simple yet sufficient model to describe the plateau of most-massive
stars between 1000 and 4000 $M_\odot$ and the behaviour at higher
cluster mass might be the following. The model assumes that the mass of the
most-massive star is linked to the proto cluster mass due to stellar
feedback. For a range of cluster 
masses, $M_\mathrm{ecl}$ (10 to $10^6$ $M_\odot$), cluster radii,
$R_\mathrm{ecl}$ (from 0.1 to 1.0 pc), and star-formation
efficiencies, SFE (0.3 to 0.8), the velocity dispersion, $\sigma$, is
calculated by  
\begin{equation}
\sigma = \sqrt{\frac{3\pi
    G}{2}\frac{M_\mathrm{ecl}}{R_\mathrm{ecl}~\mathrm{SFE}}}, 
\end{equation}
where $G$ is Newton's gravitational constant and SFE =
$\frac{M_\mathrm{ecl}}{(M_\mathrm{ecl} + M_\mathrm{gas})}$, with
$M_\mathrm{gas}$ being the residual gas mass in the cluster forming volume.

\begin{figure}
\begin{center}
\includegraphics[width=8cm]{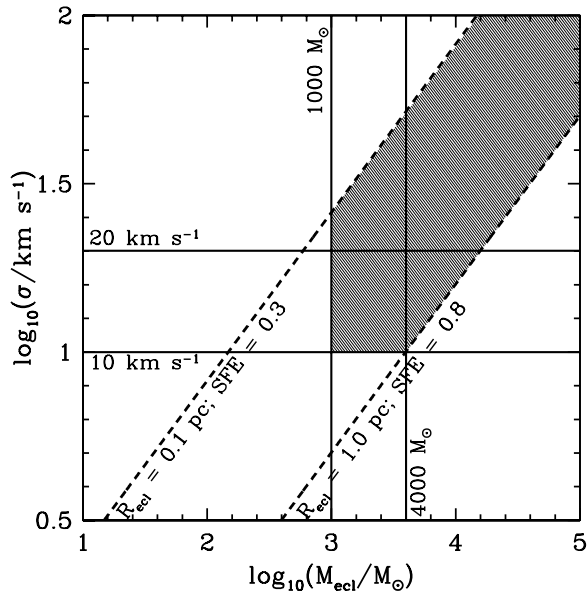}
\vspace*{-2.0cm}
\caption{The {\it dashed lines} show the dependence of the velocity
  dispersion, $\sigma$, in the cluster on $M_\mathrm{ecl}$ for a range of cluster
  radii (0.1 to 1 pc) and star-formation efficiencies (0.3 to
  0.8). The {\it horizontal lines} mark the typical range of
  velocities for ionised gas, $v_\mathrm{ion}$ (10 to 20 $\mathrm{km
    s^{-1}}$), while 
  the {\it vertical lines} are the plateau of most-massive stars for
  $M_\mathrm{ecl}$ from 1000 to 4000 $M_\odot$. In the {\it shaded
    region} are lying models for which $\sigma$ is larger
  than $v_\mathrm{ion}$ and $M_\mathrm{ecl}$ is larger than 1000 $M_\odot$.}
\label{fig:mod}
\end{center}
\end{figure}

Fig.~\ref{fig:mod} shows $\sigma$ within
the proto cluster as a function of $M_\mathrm{ecl}$. It is
compared with the typical velocity of ionised gas, $v_\mathrm{ion}$,
which is about 10 
to 20 $\mathrm{km s^{-1}}$. As is visible in Fig.~\ref{fig:mod}, $\sigma$ is
larger than $v_\mathrm{ion}$ for clusters with masses larger than a
couple of hundred $M_\odot$, regardless of the radii and SFEs. Therefore, 
it seems possible that such clusters are able to retain the ionised
gas longer - allowing the stars to accrete further mass. The fact that
$\sigma$ already overcomes $v_\mathrm{ion}$ at rather low
$M_\mathrm{ecl}$ for small $R_\mathrm{ecl}$ and low SFE
can be seen as an indication that low-mass clusters might have lower
SFEs than massive clusters.

\section{Results \& Discussion}
\label{se:disc}

We have studied the possible dependence of the mass of the
most-massive star, $m_\mathrm{max}$, on the stellar mass,
$M_\mathrm{ecl}$, of the host birth cluster. To this effect we have
significantly increased the data sample $m_\mathrm{max}(M_\mathrm{ecl})$.

Using the new spectral-type--stellar-mass conversion from \citet{MSH05}
and the here presented sample of most-massive stars in star
clusters, it has been shown here that
the observed sample divides into three sub-samples, the first being
clusters with $M_\mathrm{ecl} <$ 100 $M_\odot$, followed by clusters
between 100 and 1000 $M_\odot$ and clusters with $M_\mathrm{ecl} >$ 1000
$M_\odot$. Furthermore, there is a plateau of constant $m_\mathrm{max}$
$\approx$ 25 $M_\odot$ for clusters with masses between 1000 and 4000
$M_\odot$. 

\begin{itemize}
  \item $M_\mathrm{ecl} < 100 M_\odot$: The percentage of stars
    between the 1/6th and 5/6th quantiles is 89\% (83\% when taking
    the error bars into account) which is too tight for random
    sampling (66\%). Such a distribution is highly unlikely with
      a chance of only 
      0.2\% (0.1\% with errors) when calculated from a Binomial distribution.
    But the distribution around the median and the
    Wilcoxon singed rank test are compatible with random sampling at a
    significance of 2 percent. 
  \item $100 < M_\mathrm{ecl} \le 1000 M_\odot$: 77\% (70\% with
    errors) of the stars are within the 1/6th and 5/6th quantiles
    which is somewhat tighter than expected for random sampling
    (66\%). The probability of this to occur is rather high with
      8\% (13\% with errors). 
    But 87\% of all clusters are below the random-sampling
    median where only 50\% would be expected and the Wilcoxon singed
    rank test gives a very low probability ($1.9 \cdot 10^{-7}$) that
    the data are distributed symmetrically around the median.
  \item $M_\mathrm{ecl} > 1000 M_\odot$: Only 12\% of the data points
    (66\% with errors) are in the 2/3rd interval which is far below
    the expectation from random sampling (66\%). The probability
      for a random occurance of such a low number with the 2/3rd
      interval is $4 \cdot 10^{-11}$. Furthermore 97\% of
    the data points are lower than the median and the Wilcoxon singed
    rank test results in a very low probability ($2.8 \cdot 10^{-9}$)
    for a symmetric distribution, too.
\end{itemize}

The clusters in the mass range below 100 $M_\odot$ are the ones
most compatible with the hypothesis of being randomly sampled from
the IMF. This is also roughly the range of clusters studied by
\citet{MC08}. Their result, that the most-massive stars in these
clusters could be randomly drawn from a universal IMF, is therefore in
accordance with our conclusions. The difference is that
here it is shown that this assumption can not be generalised for more
massive/richer clusters. 

\begin{figure}
\begin{center}
\includegraphics[width=8cm]{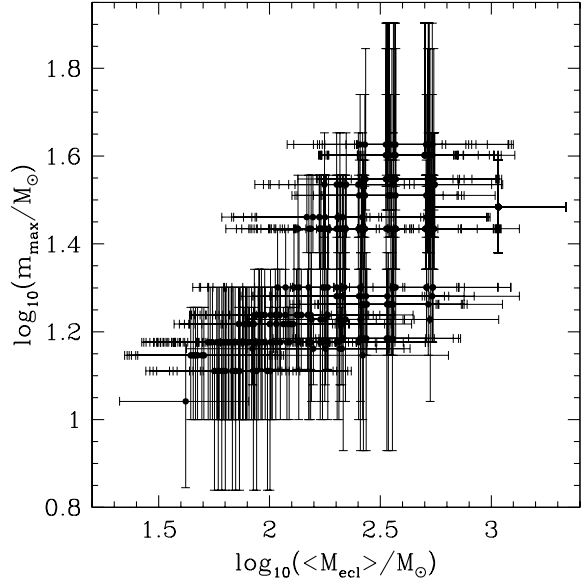}
\vspace*{-2.0cm}
\caption{The mass of the most-massive star vs the mean mass of the
  so-called ``superclusters'' constructed as in \citet{SM08}.  It
  shows whether or not the observed clusters of higher masses can be
  made by adding up large numbers of low-mass clusters. The obvious
  trend of increasing $m_\mathrm{max}$ with $<M_\mathrm{ecl}>$ is a clear
  indication against such a conclusion.}
\label{fig:selman}
\end{center}
\end{figure}

\citet{SM08} argue that the claim reached by \citet{WK05b}, that there
exists a $m_\mathrm{max}(M_\mathrm{ecl})$-relation, is due to a
size-of-sample effect in the data used by \citet{WK05b}. We now apply
their analyses to our new data set. In
appendix~A of their paper they use a method of adding up some clusters
of the \citet{WK05b} sample to so-called ``superclusters'' of the same
mass as NGC 6530 (about 1000 $M_\odot$ in the new sample presented
here). By comparing the mean mass of the synthetic superclusters with
the most-massive star of the component clusters, they show that there
is no trend for the most-massive stars to be more massive with cluster
mass. Here we repeat the same method with our new 
sample of clusters. All possible combinations to reach the mass of NGC
6530 from within the sample are used and the most-massive star is plotted
over the mean cluster mass, $<M_\mathrm{ecl}>$, in
Fig.~\ref{fig:selman}. As is seen in the figure the mass of the
most-massive star increases with $<M_\mathrm{ecl}>$. The
\citet{SM08} explanation for the \citet{WK05b} result therefore
fails for the new sample.

These results strongly suggest an underlying physical
$m_\mathrm{max}$--$M_\mathrm{ecl}$-relation. They contradict the
hypothesis that star clusters are populated with stars by random
sampling from the IMF. Only when taking into account the full range of
the error bars {\it and} a very unlikely low fundamental upper mass
limit of $m_\mathrm{max *}$ = 50 $M_\odot$ would the complete sample
mostly agree with random sampling. But in such a case no stars above 50
$M_\odot$ would exist, a result clearly disproved by the dynamical
mass measurements for the massive stars in Westerlund 2 and NGC
3603 (see Tab.~\ref{tab:dyn}).

The general trend of the most-massive star with cluster mass and the
observed plateau between the two cluster mass regimes is therefore
most likely a general result of the star-formation process within
cluster-forming molecular cloud cores. Several different mechanisms
might be responsible for the non-random behaviour of the formation of
the most-massive star in star clusters. One such model is explored in
\S~\ref{sub:mod}, where the velocity dispersion within the
cluster-forming cloud core is used as a measure for the binding energy
of the cloud, and is compared with typical velocities of ionised gas 
which acts as a proxy for the radiative feedback of the stars. This
simple model is already in qualitative agreement with the data, but more
detailed studies of how the radiative and mechanical feedback of
massive stars scales differently than the binding energy are
needed. This may result in a critical $M_\mathrm{ecl}$ limit at which
the one dominates over the other. 

Another possible explanation for the 
existence of an $m_\mathrm{max}(M_\mathrm{ecl})$-relation might be given
by dry mergers. In this scenario massive stars form in smaller
sub-clusters which are quickly evacuated by their feedback and these
sub-clusters then merge nearly gas-free, allowing only for very little
additional accretion, ie mass growth. Only for initially very massive
giant molecular clouds, more gas might be accreted during and after
the merging of the sub-clusters.

The interesting split of the massive clusters into a tight and a loose
subset by \citet[][see \S~\ref{sub:pfalz}]{Pa09} can be used as an
additional constraint on the $m_\mathrm{max}$--$M_\mathrm{ecl}$-relation. 
The loose cluster stars form predominately by free-fall collapse
of dense cores with little or no further gas accretion into the
cluster. But for the tight (high-density) clusters
cluster-potential-assisted-accretion is possible which allows for more
massive stars to form in these objects. Also stellar collisions,
mergers and competitive accretion might play a role in these dense clusters.

A more detailed study of the possible mechanisms to explain the here
presented observational evidence for a physical relation between
$m_\mathrm{max}$ and $M_\mathrm{ecl}$ will be presented in a follow-on
paper.

As the high-mass regime is most important for the question
whether the integrated IMF of a galaxy is similar to the IMF 
derived locally on star cluster scales or not, this discardation of 
random sampling naturally leads the IGIMF being steeper than expected
from individual star clusters. Since the majority of stars seem to
form in star 
clusters but also these clusters are distributed according to a mass
function which is dominated by lower mass clusters, the apparent
non-randomness of these clusters lead to fewer OB stars per star in a
galaxy than expected from random sampling\footnote{The IGIMF concept
is discussed in great detail in \citet{KW03} and \citet{WK05a} and
subsequent papers.}.

\section*{Acknowledgements}
We thank Jan Pflamm-Altenburg and Thomas Maschberger for
several lengthy discussions on statistical methods. We also thank
Vasili Gvaramadze for pointing out the work of \citet{MSH05} on the
re-calibration of the masses of massive stars and Nick Moekel for
further discussions. This work made use of the Webda and the Simbad
web based databases. This work was financially supported by the
Chilean FONDECYT grant 3060096 and the CONSTELLATION European
Commission Marie Curie Research Training Network (MRTN-CT-2006-035890).

\begin{appendix}

\section{The stellar initial mass function}
\label{app:IMF}
The following multi-component power-law IMF is used throughout the paper:

{\small
\begin{equation}
\xi(m) = k \left\{\begin{array}{ll}
k^{'} \left(\frac{m}{m_{\rm H}} \right)^{-\alpha_{0}}&\hspace{-0.25cm},m_{\rm
  low} \le m < m_{\rm H},\\
\left(\frac{m}{m_{\rm H}} \right)^{-\alpha_{1}}&\hspace{-0.25cm},m_{\rm
  H} \le m < m_{0},\\
\left(\frac{m_{0}}{m_{\rm H}} \right)^{-\alpha_{1}}
  \left(\frac{m}{m_{0}} \right)^{-\alpha_{2}}&\hspace{-0.25cm},m_{0}
  \le m < m_{1},\\ 
\left(\frac{m_{0}}{m_{\rm H}} \right)^{-\alpha_{1}}
    \left(\frac{m_{1}}{m_{0}} \right)^{-\alpha_{2}}
    \left(\frac{m}{m_{1}} \right)^{-\alpha_{3}}&\hspace{-0.25cm},m_{1}
    \le m < m_{\rm max},\\ 
\end{array} \right. 
\label{eq:4pow}
\end{equation}
\noindent with exponents
\begin{equation}
          \begin{array}{l@{\quad\quad,\quad}l}
\alpha_0 = +0.30&0.01 \le m/{M}_\odot < 0.08,\\
\alpha_1 = +1.30&0.08 \le m/{M}_\odot < 0.50,\\
\alpha_2 = +2.35&0.50 \le m/{M}_\odot < 1.00,\\
\alpha_3 = +2.35&1.00 \le m/{M}_\odot.< m_{\rm max}\\
          \end{array}
\label{eq:imf}
\end{equation}}
\noindent where $dN = \xi(m)\,dm$ is the number of stars in the mass
interval $m$ to $m + dm$. The exponents $\alpha_{\rm i}$ represent the
standard or canonical IMF and have been corrected for unresolved
multiple systems \citep{Kr01,Kr02,TK07,TK08,WK07c}. The advantage
of such a multi-part power-law description is the easy integrability
and, more importantly, that {\it different parts of the IMF can be
changed readily without affecting other parts}. Note that this form is
a two-part power-law in the stellar regime, and that brown dwarfs
contribute about 4 per cent by mass only and need to be treated as a
separate population such that the IMF has a discontinuity near $m_{\rm
H}$ = 0.08 $M_\odot$ with $k^{'} \sim \frac{1}{3}$
\citep{KBD03,TK07,TK08}. A log-normal form below 1 $M_{\odot}$ with a
power-law extension to high masses was suggested by \citet{Ch03} but
is indistinguishable from the canonical form \citep{DHK08} and does
not cater for the discontinuity. The canonical IMF is today understood
to be an invariant Salpeter/Massey power-law slope
\citep{Sal55,Mass03} above $0.5\, M_\odot$, being independent of the
cluster density and metallicity for metallicities $Z \simgreat 0.002$ 
\citep{MH98,SND00,SND02,PaZa01,Mass98,Mass02,Mass03,La02,La02b,WGH02,BMK03,PBK04,PAK06}.

The basic assumption underlying our approach is the notion that all stars
in every cluster are drawn from this same universal parent IMF, which is
consistent with observational evidence \citep{Elme99,Kr01}.

It should be noted here that, while not indicated in eq.~\ref{eq:imf},
there is evidence of a maximal mass for stars \citep[$m_\mathrm{max}
\le m_\mathrm{max *}\,\approx\,150\,M_{\odot}$,][]{WK04}, a result
confirmed by several independent studies \citep{OC05,Fi05,Ko06}.

\section{The cluster sample}
\label{app:data}

\begin{landscape}
\begin{table*}
\caption{\label{tab:clustersold} Literature data for the empirical
cluster masses ($M_\mathrm{ecl}$), maximal star masses
($m_\mathrm{max, obs}$) within these clusters, cluster ages (age),
 distances (D), the numbers of stars above or within certain mass
 limits (in $M_\odot$), the name and the spectral type of the
 most-massive star and references for the data.}
\begin{tabular}{ccccccccccccc}
\hline
Designation&$M_{\rm ecl}$&$m_{\rm max~obs~old}$&$m_\mathrm{max~obs~new}$&age & D& \# of stars&Id
$m_{\rm  max}$&Sp Type&Ref.\\
&[$M_{\odot}$]&[$M_{\odot}$]&[$M_{\odot}$]&[Myr]&[pc]&&&&\\
\hline
IRAS 05274+3345&14 -7/+15&7.0 $\pm$ 2.5&&1.0&1800&15 $>$ 0.24&-&B2&(1)\\
Mol 139&16 $\pm$ 8&2.9 $\pm$ 2.0&&$<$1&7300&-&-&-&(2)\\
Mol 143&21 $\pm$ 10&3.1 $\pm$ 2.0&&$<$1&5000&-&-&-&(2)\\
IRAS 06308+0402&24 -13/+25&11.0 $\pm$ 4.0&&1.0&1600&16 $>$ 0.37&-&B0.5&(1)\\
VV Ser&25 -13/+27&3.3 $\pm$ 1.0&&0.6&440&24 $>$ 0.3&VV Ser&B9e&(3)\\
VY Mon&28 -15/+29&4.1 $\pm$ 1.0&&0.1&800&26 $>$ 0.3&VY Mon&B8e&(3)\\
Mol 8A&30 $\pm$ 15&3.8 $\pm$ 2.0&&$<$1&11500&-&-&-&(2)\\
IRAS 05377+3548&30 -15/+32&9.5 $\pm$ 2.5&&1.0&1800&31 $>$ 0.24&-&B1&(1)\\
Ser SVS2&31 -16/+31&2.2 $\pm$ 0.2&&2.0&259 $\pm$ 37&50 $>$ 0.17&BD
01$^\circ$ 3689&A0&(4)\\
Tau-Aur&31 -18/+36&7.0 $\pm$ 2.5&&1-2&140&123 $>$ 0.02&HK Tau/G1&B2&(5)\\
IRAS 05553+1631&31 -16/+33&9.5 $\pm$ 2.5&&1.0&2000&28 $>$ 0.28&-&B1&(1)\\
IRAS 05490+2658&33 -17/+36&7.0 $\pm$ 2.5&&1.0&2100&30 $>$ 0.29&-&B2&(1)\\
IRAS 03064+5638&33 -17/+36&11.0 $\pm$ 4.0&&1.0&2200&27 $>$ 0.31&-&B0.5&(1)\\
IRAS 06155+2319&34 -18/+35&9.5 $\pm$ 2.5&&1.0&1600&38 $>$ 0.21&-&B1&(1)\\
Mol 50&36 $\pm$ 18&3.5 $\pm$ 2.0&&$<$1&4900&-&-&-&(2)\\
Mol 11&47 $\pm$ 20&3.8 $\pm$ 2.0&&$<$1&2100&-&-&-&(2)\\
IRAS 06058+2138&51 -27/+54&7.0 $\pm$ 2.5&&1.0&2000&26 $>$ 0.49&-&B2&(1)\\
NGC 2023&55 -28/+58&8.0 $\pm$ 2.0&&3.0&400&21 $>$ 0.6&HD 37903&B1.5V&(6)\\ 
Mol 3&61 $\pm$ 20&3.7 $\pm$ 2.0&&$<$1&2170&-&-&-&(2)\\
Mol 160&63 $\pm$ 20&4.3 $\pm$ 2.0&&$<$1&5000&-&-&-&(2)\\
NGC 7129&63 -33/+104&9.2 $\pm$ 3.0&&0.1&1000&53 $>$ 0.3 / 3 $>$ 3&BD
65$^\circ$ 1637&B3e&(7)\\
IRAS 06068+2030&67 -35/+70&11.0 $\pm$ 4.0&&1.0&2000&59 $>$ 0.28&-&B0.5&(1)\\
IRAS 00494+5617&71 -37/+74&9.5 $\pm$ 2.5&&1.0&2200&58 $>$ 0.31&-&B1&(1)\\
IRAS 05197+3355&72 -38/+75&11.0 $\pm$ 4.0&&1.0&3200&34 $>$ 0.50&-&B0.5&(1)\\
Cha I&80 -46/+91&5.0 $\pm$ 3.0&&2.0&170 $\pm$ 10&237 $>$ 0.04&HD
96675&B6IV/V&(8)\\
V921 Sco&71 -36/+429&14.0 $\pm$ 4.0&&0.1-1&800&33 $>$ 0.5&V921 Sco&B0e&(7)\\
IRAS 05375+3540&73 -38/+78&11.0 $\pm$ 4.0&&1.0&1800&74 $>$ 0.24&-&B0.5&(1)\\
IRAS 02593+6016&78 -41/+81&15.0 $\pm$ 5.0&&1.0&2200&61 $>$ 0.31&-&B0&(1)\\
Mol 103&80 $\pm$ 20&4.0 $\pm$ 2.0&&$<$1&4100&-&-&-&(2)\\
NGC 2071&80 -44/+89&4.0 $\pm$ 2.0&&1.0&400&105 $>$ 0.2&V1380 Ori&B5&(9)\\
MWC 297&85 $\pm$60&8.3 -1.3/+13.7&8.3 -1.3/+13.7&0.1-1&250 or 450&24
$>$ 0.3&-&B1.5V or O9e&(3)\\
IC 348&89 -46/+92&6.0 $\pm$ 1.0&&1.3&310&173 $>$ 0.1&BD
31$^{\circ}$643&B5V&(10)\\
BD 40$^\circ$ 4124&90 -49/+106&12.9 -6.0/+ 2.0&&0.1-6&1000&74 $>$ 0.3
/ 3 $>$ 3&BD 40$^\circ$ 4124&B2e&(3)\\
$\rho$ Oph&91 -46/+93&8.0 $\pm$ 1.0&&0.1-1&130 $\pm$ 20&78 $>$
0.3&$\rho$ Oph&BIV&(11)\\
IRAS 06056+2131&92 -49/+97&7.0 $\pm$ 2.5&&1.0&2000&85 $>$ 0.28&-&B2&(1)\\
IRAS 05100+3723&98 -51/+103&15.0 $\pm$ 5.0&&1.0&2600&63 $>$ 0.38&-&B0&(1)\\
R CrA&105 -55/+114&4.0 $\pm$ 2.0&&1.0&130&55 $>$ 0.5&R CrA&A5eII&(12)\\ 
NGC 1333&105 -54/+111&5.0 $\pm$ 1.0&&1-3&250&134 $>$ 0.2&SSV 13&-&(13)\\ 
Mol 28&105 $\pm$ 20&9.9 $\pm$ 2.0&&$<$1&4500&-&-&-&(2)\\
IRAS 02575+6017&111 -57/+116&9.5 $\pm$ 2.5&&1.0&2200&91 $>$ 0.31&-&B1&(1)\\
W40&144 -80/+576 100& 10.0 $\pm$ 5.0&&1-2&600&3 $>$ 4&IRS 2a&-&(14)\\
$\sigma$ Ori&150 -76/+155&20.0 $\pm$ 4.0&17.3 -4.3/+4.7&2.5&360 $\pm$
60&140 $\pm$ 10 (0.2-1.0)&$\sigma$ Ori A&O9-9.5V&(15)\\
NGC 2068&151 -86/+169&5.0 $\pm$ 3.0&&1.0&400&192 $>$ 0.2&HD 38563A&B4V&(16)\\ 
NGC 2384&189 -95/+192&16.5 $\pm$ 1.5&&1.0&2100&7 $>$ 3&HD 58509&B0.5III&(17)\\\hline
\end{tabular}
\end{table*}
\end{landscape}

\begin{landscape}
\begin{table*}
\begin{tabular}{ccccccccccccc}
\hline
Designation&$M_{\rm ecl}$&$m_\mathrm{max~obs~old}$&$m_\mathrm{max~obs~new}$&age 
&D&\# of stars&Id $m_{\rm max}$&Sp Type&Ref.\\
&[$M_{\odot}$]&[$M_{\odot}$]&[$M_{\odot}$]&[Myr]&[pc]&&&&\\
\hline
Mon R2&225 -117/+236&15.0 $\pm$ 5.0&&0-3&830 $\pm$ 50&309 $>$ 0.15&IRS
1SW&B0&(18)\\
IRAS 06073+1249&239 -120/+242&11.0 $\pm$ 4.0&&1.0&4800&25 $>$ 1.47&-&B0.5&(1)\\
Trumpler 24&251 -131/+291&14.5 $\pm$ 2.5&&1.0&1140&4 $>$ 5&GSC
7872-1609&WN&(19)\\
IC 5146&293 -226/+305&14.0 $\pm$ 4.0&&1.0&900&238 $>$ 0.3 / 5 $>$ 3&BD
46$^\circ$ 3474&B0e&(20)\\
HD 52266&400 $\pm$ 350&28.0 $\pm$ 3.5&20.0 -5.0/+7.0&$<$3.0&1700 $\pm$
1000&4 $\pm$ 2 $>$ 4 &HD 52266&O8-9V&(21)\\
HD 57682&400 $\pm$ 350&28.0 $\pm$ 3.5&20.0 -5.0/+7.0&$<$3.0&?&4 $\pm$
5 $>$ 4&HD 57682&O8-9V&(21)\\
Alicante 5&461 -234/+516&12.0 $\pm$ 4.0&&$<$3.0&3600 +600/-400&22 $>$
2.5&A47&B0.7V&(22)\\
Cep OB3b&485 -243/+497&37.7 $\pm$ 5.0&27.2 -5.2/+8.8&3.0&800 $\pm$
100&12 $>$ 4&HD 217086&O7Vn&(23)\\
HD 153426&500 $\pm$ 350&40.0 $\pm$ 6.5&28.9 -6.9/+7.1&$<$3.0&?&5 $\pm$
4 $>$ 4&HD 153426&O6.5-7V&(21)\\
NGC 2264&525 -267/+537&25.0 $\pm$ 5.0&27.2 -5.2/+8.8&3.0&760 $\pm$
100&1000 $>$ 0.08&S Mon / HD 47839&O7Ve&(24)\\
Sh2-294&525 -267/+540&12.5 $\pm$ 2.5&&4.0&3200&155 $>$
0.7&S294B0.5V&B0.5V&(25)\\
RCW 116B&536 -276/557&21.0 $\pm$ 5.0&16.9 -5.9/+7.1&2.5&1100&102 $>$
0.95&-&B1V-O8V&(26)\\
NGC 6383&561 -281/+563&37.7 $\pm$ 5.0&27.2 -5.2/+8.8&2.0&1300 $\pm$
100&21 $>$ 3&HD 159176&O7V + O7V&(27)\\
Alicante 1&577 -290/+583&45.0 $\pm$ 5.0&34.3 -8.3/+10.7&2-3&4000 $\pm$
400&38 $>$ 2.0&BD 56$^\circ$ 864&O6V&(28)\\
HD 52533&621 -417/+1077&26.7 $\pm$ 3.0&19.1 -4.1/+2.9&$<$3.0&?&15 $\pm$
5 $>$ 4&HD 52533&O8.5-9V&(21)\\
Sh2-128&666 -342/+736&37.7 $\pm$ 5.0&27.2 -5.2/+8.8&2.0&9400&7 $>$
7&-& O7V&(29)\\
NGC 2024&690 -350/+706&20.0 $\pm$ 4.0&15.3 -6.8/+8.7&0.5&400&309 $>$
0.5&IRS 2b&O8V - B2V&(30)\\
HD 195592&725 -364/+757&40.0 $\pm$10.0&32.4 -6.4/+6.6&$<$3.0&?&18
$\pm$ 3 $>$ 4&HD 195592&O6-6.5V&(21)\\
Sh2-173&748 -395/+901&25.4 $\pm$ 5.0&18.3 -4.3/+3.7&0.6-1.0&2500 $\pm$
500&7 $>$ 7&BD 60$^\circ$ 39&O9V&(31)\\
DBSB 48&792 -416/+1126&56.6 $\pm$ 15.0&42.3 -8.3/+12.7&1.1&5000 $\pm$
700&5 $>$ 10&-&O5V&(32)\\
NGC 2362&809 -409/+823&43.0 $\pm$ 7.0&35.3 -5.3/+34.7&3.0&1390 $\pm$
200&353 $>$ 0.5&$\tau$ CMa&O9Ib&(33)\\
Pismis 11&896 -448/+938&40.0 -0.0/+40.0&&3-5&3600 +600/-400&43 $>$
2.5&HD 80077&B2Ia&(22)\\
NGC 6530&1075 -545/+1097&56.6 $\pm$ 11.0&30.5 -6.5/+8.5&2.3&1350 $\pm$
200&620 $>$ 0.4&HD 165052&O6.5V&(34)\\
FSR 1530&1410 -707/+1581&30.0 $\pm$ 15.0&&$<$4.0&2750 $\pm$ 750&35 $>$
4&[M81]I-296&-&(35)\\
Berkeley 86&1440 -730/+1470&40.0 $\pm$ 8.0&21.7 -4.7/+7.3&3-4&1700&340
$>$ 0.8&HD 193595&O8V(f)&(36)\\
NGC 637&1682 -854/+1726&30.8 $\pm$ 15.0&21.7 -6.7/+11.3&4.0&2160&583
$M_{\odot}$ $>$ 1.6&-&$\approx$ O8&(37)\\
W5Wb&1734 -874/+1757&34.1 $\pm$ 5.0&24.1 -4.1/+8.9&2.0&2000&300 $>$
1&BD 60$^\circ$ 586&O7.5V&(38)\\
Stock 16&1857 -955/+2045&43.0 $\pm$ 10.0&32.1 -7.1/+7.9&4.0&1650&16
$>$ 8&HD 115454&O7.5III&(39)\\
ONC&2124 -1078/+2175&35.8$^a$ $\pm$ 7.2&34.3 - 5.7/+10.7&$<$1.0&414
$\pm$ 7&3500 (0.1 - 30)&$\Theta$ Orionis C1&O6Vpe&(40)\\
RCW 38&2251 -1132/+2276&50.4 $\pm$ 10.0&37.7 -7.7/+12.3&$<$1.0&1700&2000
$>$ 0.25&IRS 2&O5.5V&(41)\\
Bochum 2&2284 -1302/+3523&37.7 $\pm$ 4.0&27.2 -5.2/+8.8&2-4&2700&4 $>$
16&BD 00$^\circ$ 1617B&O7V&(42)\\
Berkeley 59&2310 -1168/+2417&37.7 $\pm$ 4.0&27.2 -5.2/+8.8&2.0&1000&41
$>$ 5&BD 66$^\circ$ 1675&O7V&(43)\\
IC 1590&2376 -1245/+2799&41.0 $\pm$ 5.2&30.5 -6.5/+8.5&3.5&2900&14 $>$
10&BD 55$^\circ$ 191&O6.5V&(44)\\
W5E&2614 -1323/+2667&37.7 $\pm$ 5.0&27.2 -5.2/+8.8&2.0&2000&400 - 500
$>$ 1&HD 18326&O7V&(38)\\
W5Wa&2651 -1338/+2690&52.0 $\pm$ 10.0&37.3 -6.3/+7.7&2.0&2000&400 - 500
$>$ 1&HD 17505&O6.5III&(38)\\
NGC 1931&3128 -1564/+3163&30.8 $\pm$ 15.0&21.7 -6.7/+11.4&4.0&3086&848
$M_{\odot}$ $>$ 2.39&-&$\approx$ O8&(37)\\
LH 118&3746 -1918/+4077&56.6 $\pm$ 7.0&42.3 -8.3/+12.7&3.0&48500&28 $>$
9&LH 118-241&O5V&(45)\\
NGC 2103&3853 -1937/+3905&80.0 $\pm$ 20.0&89.3 -24.3/+30.7&1.0&48500&26
$>$ 10&Sk -71$^\circ$ 51&O2V((f$^{\ast}$))&(46)\\
NGC 7380& 4527 -2290/+4611&47.8 $\pm$ 5.0&36.0 -6.0/+9.0&2.0&3700&42 (6
- 12)&HD 215835&O5.5-6V((f))&(47)\\
NGC 6231&4595 -2312/+4676&59.0 $\pm$ 10.0&51.9 -20.9/+18.1&1.0&1600&51
$>$ 7&HD 152248&O7Ib&(48)\\
RCW 106&4681 -2375/+4871&45.7 $\pm$ 10.0&34.1 -8.1/+10.9&2.5&1100&41 $>$
8&-&O5.5-6.5V&(49)\\
NGC 6823&4983 -2584/+5685&37.7 $\pm$ 4.0&27.2 -5.2/+8.8&2-4&1900&42 $>$
8&BD 22$^\circ$ 3782&O7V(f)&(50)\\
RCW 121&5323 -2671/+5390&42.0 $\pm$ 4.0&38.3 -8.3/+11.7&4.2&1600&96 $>$
5&-&O6V-O5V&(51)\\
NGC 2244&5946 -3029/+6102&68.9 $\pm$ 14.0&54.1 -9.1/+10.9&1.9&1500&54
(6 - 12)&HD 46223&O4V(f)&(52)\\
NGC 2122&6764 -3416/+6960&74.7 $\pm$ 10.0&58.8 -13.8/+11.2&3.0&48500&52
$>$ 9&HD 270145 & O6I(f)&(53)\\
$[$OBS 2003$]$ 179& 7000 $\pm$ 3500&50.0 $\pm$ 20.0&70.0 $\pm$ 10.0&2-5&7900&10 $>$
16&Obj 4&Ofpe/WN9&(54)\\
Westerlund 2&8845 -4456/+9009&82.7$^{a}$ $\pm$ 5.5&121.0
-43.8/+29.0&1-2&8000&29 $>$ 16.5&WR20a A&WN6ha&(55)\\\hline
\end{tabular}

$^a$ Dynamical mass estimates from binary orbits exist for these stars.\\

\end{table*}
\end{landscape}

\begin{landscape}
\begin{table*}
\begin{tabular}{ccccccccccccc}
\hline
Designation&$M_{\rm ecl}$&$m_\mathrm{max~obs~old}$&$m_\mathrm{max~obs~new}$& 
age&D&\# of stars&Id $m_{\rm max}$&Sp Type&Ref.\\
&[$M_{\odot}$]&[$M_{\odot}$]&[$M_{\odot}$]&[Myr]&[pc]&&&&\\
\hline
RCW 95&9670 -4840/+9720&87.6 $\pm$ 14.0&64.6 -9.6/+15.4& 
1.5&2400&136 $>$ 6&-&O3V&(56)\\
IC 1805&10885 -5528/+11137&75.6 $\pm$ 10.0&56.5 -6.5/+8.5 &2.0&2350&99
(6 - 12)&HD 15558&O4-5III(f)&(47)\\
NGC 6357&11978 -6430/+11979&115.9 $\pm$ 30.0&79.8 -14.8/+15.2&
1.0&2560&38 $>$ 16.5&HDE 319718A&O3If&(57)\\
NGC 3603&$1.3 \cdot 10^{4}$ $\pm$ 3000&150.0 $\pm$ 50.0&121.0 -41.0/+29.0& 
0.7&6000&-&NGC 3603-B&WN6ha&(58)\\
Trumpler 14/16&17890 -8945/+18676&150.0 $\pm$ 50.0& 99.8 -39.8/+50.2& 
1.7&2500&64 $>$ 16&$\eta$ Carina &LBV&(59)\\
NGC 6611&25310 -12659/+25503&85.0 $\pm$ 15.0&60.4 -5.6/+9.6&
1.3&1800&460 $>$ 5&HD 168076&O4III&(60)\\
Cyg OB2&75890 -38453/+78716&92.0 -25.0/+58.0&&2.0&
1700&8600 $>$ 1.3&Cyg OB2-12&B8Ia&(61)\\
Arches&77225 -39250/+77225&120.0 -0.0/+30.0& 111.0 -41.0/+39.0&
2.5&7620&196 $>$ 20.1&N4&WN7-8h&(62)\\
R 136&222912 -112104/+224426&145.0 $\pm$ 15.0&125.4 -45.4/+24.6& 
1-2&48500&8000 $>$ 3&R136a1&O2If$^\ast$/WN4.5&(63)\\
\hline
\end{tabular}

$^a$ Dynamical mass estimates from binary orbits exist for these stars.\\

\end{table*}
\end{landscape}

\begin{table}
\caption{\label{tab:ref} References for Table~\ref{tab:clustersold}.}
1:~\citet{CSS90,CSS93},
2:~\citet{FMT09},
3:~\citet{TPP97,TPN98,TPN99,WL07},
4:~\citet{KOB04},
5:~\citet{CK79,BLH02},
6:~\citet{Se83,DLG90,LDE91},
7:~\citet{GMM04,WL07},
8:~\citet{Lu08},
9:~\citet{LDE91},
10:~\citet{PZ01,LL03},
11:~\citet{WLY89,La03,WGA08},
12:~\citet{NF08},
13:~\citet{AC03,GFT02,GMM08},
14:~\citet{SBC85,RR08},
15:~\citet{SWW04,BHM08},
16:~\citet{Se83,LDE91},
17:~\citet{PBM89},
18:~\citet{CMD97,PBS02},
19:~\citet{HW84,PBM89},
20:~\citet{Wa59,FO84,PBM89,WL07,MNL07},
21:~\citet{DTP04,DTP05},
22:~\citet{MN08},
23:~\citet{NF99,PNJ03,MNL07},
24:~\citet{SBC04,MNL07,Da08},
25:~\citet{YDD08},
26:~\citet{R07},
27:~\citet{RDG03,PNZ07,RD08},
28:~\citet{NM08},
29:~\citet{BT03},
30:~\citet{LDE91,HLL00,SWW04},
31:~\citet{CRO09},
32:~\citet{OBB08},
33:~\citet{MNL07,DH07},
34:~\citet{PDM04,DFM04,DPM06,MNL07,CGZ07},
35:~\citet{FMS08},
36:~\citet{MJD95,VRC99},
37:~\citet{HHS08},
38:~\citet{KAG08},
39:~\citet{T85,PBM89},
40:~\citet{HH98,HSC98,MRF07,KWB08},
41:~\citet{WSB06,WBV08},
42:~\citet{Da99,LPS02},
43:~\citet{BW59,Mc68,PSO08},
44:~\citet{W73,GT97,LPS02},
45:~\citet{MSG89},
46:~\citet{MHW05},
47:~\citet{WSD07},
48:~\citet{GM01,SGR06,SRS07,SGN07},
49:~\citet{RAL03,R07},
50:~\citet{PKK00,LPS02},
51:~\citet{RA06,R07},
52:~\citet{MJD95,PS02,CGZ07,WSD07,BB09},
53:~\citet{SM86,GW87,MSG89,NG04},
54:~\citet{BIH08},
55:~\citet{BSU04,NRM08},
56:~\citet{RA04b,R07},
57:~\citet{BTR04,MWM07,WTF07},
58:~\citet{HY07,HEM07,SMS08},
59:~\citet{PGH93,MJ93,MJD95,NWW04,OC05,AAV07,SCO07,OBB08},
60:~\citet{BSS06,WSD07},
61:~\citet{K00,MDW01,WSD07,NMH08},
62:~\citet{FNG02,MHP08},
63:~\citet{MH98,SMB99,SCC09}
\end{table}

\begin{figure}
\begin{center}
\includegraphics[width=8cm]{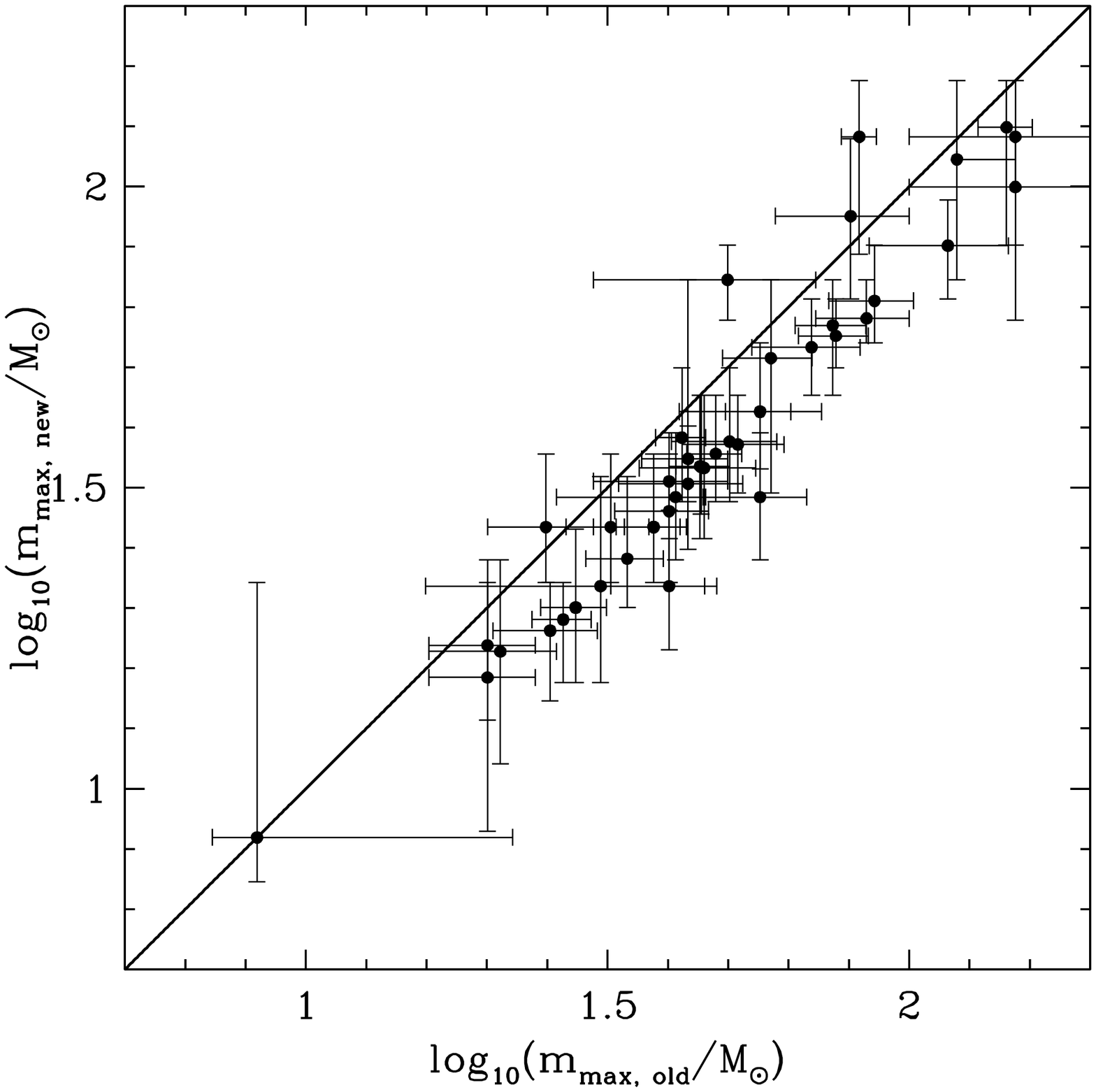}
\vspace*{-2.0cm}
\caption{The values for the O star masses from the old calibration on
  the abscissae vs the newly calibrated ones on the ordinate. The {\it
    thick solid line} indicates were both values would be the same.}
\label{fig:old_vs_new}
\end{center}
\end{figure}

\section{Notes on individual star clusters}
\label{app:notes}
\subsection{MWC 297}
Depending on the uncertain distance the spectral type of the brightest
star is either O9e \citep[450 pc,][]{HSV92} or B1.5V \citep[250
  pc,][]{DBH97}. \citet{WL07} assume the O9e spectral type and assign
the star a mass of 26.5 $M_\odot$. This mass estimate is rather odd as
older literature gives 22 \citep{HHC97} or 25.4 \citep{VGS96}
$M_\odot$ for a O9 star and the new \citet{MSH05} calibration yields 18
$M_\odot$. Here we assume the newer distance of 250 pc and therefore
the spectral type B1.5V but include the mass of an O9V star as an
upper limit (22 $M_\odot$ for the old calibration and 18 $M_\odot$ for
the new one).

\subsection{Pismis 11}
It is not certain whether HD 80077 is part of this star cluster or not
\citep{MN08}. If it is it would be one of the brightest stars in the
MW. As it is an evolved star (B2Ia) the precise initial mass in not
known, only that it should be above 40 $M_\odot$.

\subsection{NGC 2244}
\citet{BB09} give a mass of 625 $M_\odot$ for this cluster from
fitting a King profile to the MS and PMS stars they found. But when
using the number of B0 to B3 stars (54 between 6 and 12 $M_\odot$
$M_\odot$) from \citet{WSD07}, an IMF extrapolation yields a cluster
mass of about 6000 $M_\odot$.  A reason for this discrepancy is
currently not known.

\subsection{$\gamma$ Velorum Cluster}
\citet{JNW09} studied a group of several hundred PMS stars around the
massive binary $\gamma^2$ Vel (WC8 + O7.5V) in the Vela OB2
association. They concluded that it is a young ($\approx$ 7 Myr)
low-mass (250 - 360 $M_\odot$) cluster with a very massive star (initial
mass for the WC8 star $>$ 40 $M_\odot$). This object has not been included
because of its age and the possible mass loss from the cluster due to
gas expulsion.

\subsection{Trumpler 14/16}
Trumpler 14 and Trumpler 16 are believed to be one single cluster
optically divided by a large patch of gas and dust \citep{OC05}. Some
studies do not consider $\eta$ Carina as part of this cluster. If
$\eta$ Car is not part of the cluster then the O3If$^{\ast}$ star HD
93129AB (67$^+$ $M_\odot$) would be the most-massive star in the cluster.

\subsection{Cyg OB2}
The star Cyg OB2-12 is an evolved star (B8Ia) and is considered to be one
of or maybe even the brightest star in the MW. \citet{NMH08} estimate
its initial mass with 92 $M_\odot$. The region contains 8600 stars
earlier than F3, some 2600 OB stars and around 120 O stars and has
been described by \citet{K00} as a possible young globular cluster
within the MW disk.

\subsection{R 136}
There is some debate in the literature whether or not extremely
massive stars ($>$ 80 $M_\odot$) might or might not look like
Wolf-Rayet stars even while they still burn hydrogen in their cores
\citep{Cr07}. Therefore, the spectral type of the most-massive star
(R136a1) might be either O2If$^\ast$ or WN4.5. But this has little
consequences for the mass estimate for this star as it is based on a
detailed spectral analysis \citep{MH98}. The mass estimate for the
cluster, $2.2 \cdot 10^5 M_\odot$, is for R 136 only. The cluster contains
about 8000 stars between 3 and 120 $M_\odot$ and 39 O3If$^\ast$
stars. The whole region of 30 Doradus is believed to hold a total of
$4.5 \cdot 10^5 M_\odot$ in stars \citep{BTT08}.

\subsection{Cl 1806-20}
The cluster and its most massive star, LBV 1806-20, one of the
brightest stars in the MW, have not been included because the cluster
is known to host a pulsar (SGR 1806-20) and is therefore most probably
too old for the current study. Additionally, very little is known
about its stellar population besides the LBV, the pulsar, an OB
supergiant and three Wolf-Rayet stars as this cluster is on the far
side of the Galaxy (D $\approx$ 15000 pc) and heavily dust obscured
\citep{FNG05}.

\subsection{Quintuplet}
This massive cluster \citep[$\sim 1.2 \cdot 10^4$
  $M_\odot$,][]{FKM99,LHO09} hosts the Pistol star which is one of the
brightest and most massive stars in the Galaxy (150$^+$ $M_\odot$?)
but is believed to be 4-6 Myr old and therefore too old for this study.

\subsection{Westerlund 1}
Another massive cluster \citep[$>$ 10$^4$ $M_\odot$, ][]{Bw08} with about
150 O stars. It houses an X-ray pulsar \citep{MCC06} and is therefore
not included into this study as it is too old (4-5 Myr).
\end{appendix}
\bibliography{mybiblio}
\bsp
\label{lastpage}
\end{document}